\documentclass[10pt]{article}

\usepackage{fancyhdr}
\usepackage{extramarks}
\usepackage{amsmath}
\usepackage{amsthm}
\usepackage{amsfonts}
\usepackage{siunitx}
\usepackage{tikz}
\usepackage[plain]{algorithm}
\usepackage{algpseudocode}
\usepackage{multirow}
\usepackage{booktabs}
\usepackage{palatino}
\usepackage{graphicx}
\usepackage{subfigure}
\usepackage[colorlinks,linkcolor=black,anchorcolor=black,citecolor=black,urlcolor=blue]{hyperref}
\usepackage{amsmath,bm}
\usepackage{booktabs}
\usepackage{mathtools}
\usepackage{amssymb}
\usepackage{caption}
\usepackage{capt-of}
\usepackage{mciteplus}
\usepackage{cite}
\usepackage{mathrsfs}
\usepackage[title,titletoc,toc]{appendix}
\usepackage{xr}
\usepackage{parskip}
\usetikzlibrary{automata,positioning}

\renewcommand{\part}[1]{\textbf{\large Part \Alph{partCounter}}\stepcounter{partCounter}\\}

\begin{document}

\title{Repositioning of 8565 existing drugs for COVID-19}
\author{Kaifu Gao$^1$, Duc Duy Nguyen$^1$, Jiahui Chen$^1$, Rui Wang$^1$, and Guo-Wei Wei$^{1,2,3}$ \footnote{Address correspondences to Guo-Wei Wei. E-mail:wei@math.msu.edu} \\
$^1$ Department of Mathematics,
Michigan State University, MI 48824, USA\\
$^2$  Department of Biochemistry and Molecular Biology\\
Michigan State University, MI 48824, USA \\
$^3$ Department of Electrical and Computer Engineering \\
Michigan State University, MI 48824, USA }


\maketitle

\begin{abstract}
The coronavirus disease 2019 (COVID-19) pandemic caused by severe acute respiratory syndrome coronavirus 2 (SARS-CoV-2) has infected near 5 million people and led to over 0.3 million deaths. Currently, there is no specific anti-SARS-CoV-2 medication. New drug discovery typically takes more than ten years. Drug repositioning becomes one of the most feasible approaches for combating COVID-19.  This work curates the largest available experimental dataset for SARS-CoV-2 or SARS-CoV main protease inhibitors. Based on this dataset, we develop validated machine learning models with relatively low root mean square error to screen 1553 FDA-approved drugs as well as other 7012 investigational or off-market drugs in DrugBank. We found that many existing drugs might be potentially potent to SARS-CoV-2. The druggability of many potent SARS-CoV-2 main protease inhibitors is analyzed. This work offers a foundation for further experimental studies of COVID-19 drug repositioning.

\end{abstract}


\newpage

{\setcounter{tocdepth}{4} \tableofcontents}

\renewcommand{\thepage}{{\arabic{page}}}

\section{Introduction}

Severe acute respiratory syndrome coronavirus 2 (SARS-CoV-2) outbreak in Wuhan, China, in late December 2019 and has rapidly spread around the world. By May 19, 2020, near 5M individuals were infected, and more than 300K fatalities had been reported.  Currently, there is no specific antiviral drug for this epidemic. It is worth noting that recently, an experimental drug Remdesivir, has been recognized as a promising anti-SARS-CoV-2 drug. However, the high experimental value of IC$_{50}$ (11.41 $\mu$M) \cite{jeon2020identification} might indicate its inefficiency in antiviral activities against SARS-CoV-2.  

Considering the severity of this widespread dissemination and health threats, panic patients misled by media flocked to the pharmacies for Chinese medicine herbs, which were reported to ``inhibit'' SARS-CoV-2, despite no clinical evidence supporting the claim.  Although there is also no evidence for the claimed Chloroquine's curing effect, some desperate people take it as ``prophylactic'' for COVID-19. Many researchers are engaged in developing anti-SARS-CoV-2 drugs~\cite{macintyre2020wuhan,
xu2020nelfinavir}. However, new drug discovery is a long, costly, and rigorous scientific process. A more effective approach is to search for anti-SARS-CoV-2 therapies from existing drug databases.

 Drug repositioning (also known as drug repurposing), which concerns the investigation of existing drugs for new therapeutic target indications, has emerged as a successful strategy for drug discovery due to the reduced costs and expedited approval procedures~\cite{brown2017standard,amelio2014drugsurv,jin2014toward}. Several successful examples unveil its great values in practice: Nelfinavir,  initially developed to treat the human immunodeficiency virus (HIV), is now being used for cancer treatments. Amantadine was firstly designed to treat influenza caused by type A influenza viral infection and is being used for Parkinson's disease later on~\cite{patwardhan2016innovative}. In recent years, the rapid growth of drug-related datasets, as well as open data initiatives, has led to new developments for computational drug repositioning, particularly structural-based drug repositioning (SBDR). Machine learning, network analysis, and text mining and semantic inference are three major computational approaches commonly applied in drug repositioning~\cite{li2016survey}. The rapid accumulation of genetic and structural databases \cite{bank2014electronic}, the development of low-dimensional mathematical representations of complex biomolecular structures \cite{cang2018representability,nguyen2019dg}, and the availability of advanced deep learning algorithms have made machine learning-based drug reposition a promising approach \cite{li2016survey}. Due to the urgent need for anti-SARS-CoV-2 drugs, a computational drug repositioning is one of the most feasible strategies for discovering SARS-CoV-2 drugs.

In SBDR, one needs to select one or a few effective targets. Study shows that SARS-CoV-2 genome is very close to that of the severe acute respiratory syndrome (SARS)-CoV \cite{gralinski2020return}. The sequence identities of SARS-CoV-2 3CL protease, RNA polymerase, and the spike protein with corresponding SARS-CoV proteins are  96.08\%, 96\%, and 76\%, respectively \cite{xu2020evolution}. We, therefore, hypothesize that a potent SARS 3CL protease inhibitor is also a potent SARS-CoV-2 3CL protease inhibitor. Unfortunately, there is no effective SARS therapy at present. Nevertheless, the X-ray crystal structures of both
SARS and SARS-CoV-2 3CL proteases have been reported \cite{lee2005crystal, zhang2020crystal}.
Additionally,  the binding affinities of SARS-CoV or SARS-CoV-2 3CL protease inhibitors from single-protein experiments are available in various databases or the original literature.
Moreover, the DrugBank contains about 1600 drugs approved by the U.S. Food and Drug Administration (FDA) as well as more than 7000 investigational or off-market ones \cite{wishart2018drugbank}. The aforementioned information provides a sound basis to develop an SBDR machine learning model for SARS-CoV-2 3CL protease inhibition.

In responding to the pressing need for anti-SARS-CoV-2 medications, we have carefully collected 314 bonding affinities for SARS-CoV or SARS-CoV-2 3CL protease inhibitors, which is the largest set available so far for this system. Machine learning models are built for these data points.

Unlike most earlier COVID-19 drug repositioning works, including ours \cite{nguyen2020potentially}, that did not provide a target-specific cross-validation test, we have carefully optimized our machine learning model with a 10-fold cross-validation test on SARS-CoV-2 3CL protease inhibitors. We achieve a Pearson correlation coefficient of 0.78 and a root mean square error (RMSE) of 0.80 kcal/mol, which is much better than that of similar machine learning models for standard training sets in the PDBbind database (around 1.9 kcal/mol) \cite{li2020machine,nguyen2019agl}. We systematically evaluate the binding affinities (BAs) of 1553 FDA-approved drugs as well as 7012 investigational or off-market ones in the DrugBank by our 2D-fingerprint based machine learning model. Besides, a three-dimensional (3D) pose predictor named MathPose \cite{nguyen2020mathdl} is also applied to predict the 3D binding poses. With these models, we report the top 30 potential anti-SARS-CoV-2 3CL inhibitors from the FDA-approved drugs and another top 30 from investigational or off-market ones. We also discuss the druggability of some potent inhibitors in our training set. The information provides timely guidance for the further development of anti-SARS-CoV-2 drugs.

\section{Results}

\subsection{Binding affinity prediction and ranking of 1553 FDA-approved drugs }

\begin{table}[ht]
\caption{A summary of the top 30 potential anti-SARS-CoV-2 drugs from   1553 FDA-approved drugs with their predicted binding affinities (unit: kcal/mol), $\text{IC}_{50}$ ($\mu$M), and corresponding brand names.}
\centering
\setlength\tabcolsep{3pt}
\captionsetup{margin=0.9cm}
\begin{tabular}{c|l|l|c|c} \hline
DrugID & Name & Brand name & Predicted & $\text{IC}_{50}$ \\
& & & binding affinity& \\ \hline
DB01123 & Proflavine & Bayer Pessaries, Molca, Septicide & -8.37 & 0.72 \\
DB01243 & Chloroxine & Capitrol & -8.24 & 0.89 \\
DB08998 & Demexiptiline & Deparon, Tinoran & -8.14 & 1.06 \\
DB00544 & Fluorouracil & Adrucil & -8.11 & 1.11 \\
DB03209 & Oteracil & Teysuno & -8.09 & 1.16 \\
DB13222 & Tilbroquinol & Intetrix  & -8.08 & 1.18 \\
DB01136 & Carvedilol & Coreg & -8.06 & 1.22 \\
DB01033 & Mercaptopurine & Purinethol & -8.04 & 1.26 \\
DB08903 & Bedaquiline & Sirturo & -8.02 & 1.29 \\
DB00257 & Clotrimazole & Canesten & -8.00 & 1.35 \\
DB00878 & Chlorhexidine & Betasept, Biopatch & -8.00 & 1.35 \\
DB00666 & Nafarelin & Synarel & -8.00 & 1.35 \\
DB01213 & Fomepizole & Antizol & -7.98 & 1.39 \\
DB01656 & Roflumilast & Daxas, Daliresp & -7.97 & 1.41 \\
DB00676 & Benzyl benzoate & Ascabin, Ascabiol, Ascarbin, Tenutex & -7.96 & 1.45 \\
DB06663 & Pasireotide & Signifor & -7.95 & 1.47 \\
DB08983 & Etofibrate & Lipo Merz Retard, Liposec & -7.94 & 1.48 \\
DB06791 & Lanreotide & Somatuline & -7.94 & 1.48 \\
DB00027 & Gramicidin D & Neosporin Ophthalmic & -7.94 & 1.48 \\
DB00730 & Thiabendazole & Mintezol, Tresaderm, and Arbotect & -7.93 & 1.51 \\
DB00643 & Mebendazole & Vermox, Emverm & -7.90 & 1.59 \\
DB01275 & Hydralazine & Apresoline & -7.90 & 1.60 \\
DB04920 & Clevidipine & Cleviprex & -7.89 & 1.61 \\
DB01184 & Domperidone & Motilium & -7.89 & 1.61 \\
DB00150 & Tryptophan & Tryptan, Aminomine & -7.89 & 1.63 \\
DB00724 & Imiquimod & Aldara, Zyclara & -7.88 & 1.63 \\
DB01065 & Melatonin & Bio-Melatonin, SGard & -7.87 & 1.68 \\
DB00239 & Oxiconazole & Oxistat & -7.86 & 1.69 \\
DB11071 & Phenyl salicylate & Urimax & -7.86 & 1.69 \\
DB08909 & Glycerol phenylbutyrate & Ravicti & -7.85 & 1.73 \\\hline
\end{tabular}
\label{Summary1}
\end{table}

 With the SARS-CoV-2 3CL protease as the target, we predict the binding affinities of  1553 FDA-approved drugs using our machine learning predictor. Based on these predicted affinities, the top 30 potential SARS-CoV-2 inhibitors from the FDA-approved drugs are shown in Table \ref{Summary1}. A complete list of the predicted values for 1553 FDA-approved drugs is given  in the Supporting Material.

We briefly describe the top 10 predicted potential anti-SARS-CoV-2 drugs from the FDA-approved set. The most potent one is Proflavine, an acriflavine derivative. It is a disinfectant bacteriostatic against many gram-positive bacteria. Proflavine is toxic and carcinogenic in mammals and so it is used only as a surface disinfectant or for treating superficial wounds. Under the circumstance of the SARS-CoV-2, this drug might be used to clean skin or SARS-CoV-2 contaminated materials, offering an extra layer of protection. The second drug is Chloroxine, also an antibacterial drug, which is used in infectious diarrhea, disorders of the intestinal microflora, giardiasis, inflammatory bowel disease. It is notable that this drug belongs to the same family with Chloroquine, which was once considered for anti-SARS-CoV-2. However, according to our prediction, Chloroquine is not effective for SARS-CoV-2 3CL protease inhibition (BA: -6.92 kcal/mol).
The third one, Demexiptiline, a tricyclic antidepressant,  acts primarily as a norepinephrine reuptake inhibitor. The next one, Fluorouracil, is a medication used to treat cancer. By injection into a vein, it is used for colon cancer, esophageal cancer, stomach cancer, pancreatic cancer, breast cancer, and cervical cancer. The fifth drug, Oteracil, is an adjunct to antineoplastic therapy, used to reduce the toxic side effects associated with chemotherapy. The next one, Tilbroquinol, is a medication used in the treatment of intestinal amoebiasis. The seventh drug, Carvedilol, is a medication used to treat high blood pressure, congestive heart failure, and left ventricular dysfunction. The number eight drug, Mercaptopurine, is a medication used for cancer and autoimmune diseases. Specifically, it treats acute lymphocytic leukemia, chronic myeloid leukemia, Crohn's disease, and ulcerative colitis. The next one is Bedaquiline, which is a medication used to treat active tuberculosis, specifically multi-drug-resistant tuberculosis along with other tuberculosis. The number ten drug, Clotrimazole, is an antifungal medication, which is used to treat vaginal yeast infections, oral thrush, diaper rash, pityriasis versicolor, and types of ringworm including athlete's foot and jock itch.

\begin{table}[ht!]
	\caption{A summary of top 30 potential anti-SARS-CoV-2 drugs from 7012 investigational or off-market drugs with predicted binding affinities (BAs) (unit: kcal/mol), $\text{IC}_{50}$ ($\mu$M), and corresponding trade names.}
	\centering
	\setlength\tabcolsep{3pt}
	\captionsetup{margin=0.9cm}
	\begin{tabular}{c|l|c|c} \hline
		DrugID & Name & Predicted & $\text{IC}_{50}$ \\
		& & BA & \\ \hline
		DB12903 & DEBIO-1347 &  -9.02 & 0.24 \\
		DB07959 & 3-(1H-BENZIMIDAZOL-2-YL)-1H-INDAZOLE & -9.01 & 0.24 \\
		DB07301 & 9H-CARBAZOLE &  -8.96 & 0.27 \\
		DB07620 & 2-[(2,4-DICHLORO-5-METHYLPHENYL)SULFONYL] &  -8.89 & 0.30 \\
		&-1,3-DINITRO-5-(TRIFLUOROMETHYL)BENZENE & & \\
		DB08036 & 6,7,12,13-tetrahydro-5H-indolo[2,3-a]pyrrolo[3,4-c]carbazol-5-one & -8.89 & 0.30 \\
		DB08440 & N-1,10-phenanthrolin-5-ylacetamide & -8.83 & 0.33 \\
		DB01767 & Hemi-Babim & -8.80 & 0.35 \\
		DB06828 & 5-[2-(1H-pyrrol-1-yl)ethoxy]-1H-indole & -8.73 & 0.39 \\
		DB14914 & Flortaucipir F-18 & -8.69 & 0.42 \\
		DB15033 & Flortaucipir & -8.69 & 0.42 \\
		DB13534 & Gedocarnil & -8.67 & 0.44 \\
		DB02365 & 1,10-Phenanthroline & -8.64 & 0.45 \\
		DB09473 & Indium In-111 oxyquinoline & -8.64 & 0.45 \\
		DB08512 & 6-amino-2-[(1-naphthylmethyl)amino]-3,7 & -8.60 & 0.48 \\
		& -dihydro-8H-imidazo[4,5-g]quinazolin-8-one  & &\\
		DB01876 & Bis(5-Amidino-2-Benzimidazolyl)Methanone & -8.60 & 0.49 \\
		DB07919	& 7-METHOXY-1-METHYL-9H-BETA-CARBOLINE	& -8.59	& 0.49	\\
		DB02089	& CP-526423	& -8.59	& 0.50 \\
		DB07837	& [4-(5-naphthalen-2-yl-1H-pyrrolo[2,3-b] & -8.53 & 0.55 \\
		& pyridin-3-yl)phenyl]acetic acid & & \\
		DB08073	& (2S)-1-(1H-INDOL-3-YL)-3-\{[5-(3-METHYL-1H-INDAZOL-5-YL)	& -8.53 &0.55 \\
		& PYRIDIN-3-YL]OXY\}PROPAN-2-AMINE & & \\
		DB08267	& 6-amino-4-(2-phenylethyl)-1,7-dihydro-8H-imidazo[4,5-g] &-8.52 & 0.56	\\
		& quinazolin-8-one && \\
		DB02390	& 5-Bromo-N[2-(Dimethylamino)Ethyl]-9-Aminoacridine-4 &	-8.50 & 0.57 \\
		& -Carboxamide && \\
		DB07588	& 3,4-DI-1H-INDOL-3-YL-1H-PYRROLE-2,5-DICARBOXYLIC ACID	&-8.50	&0.58	\\
		DB03213	& Bis(5-Amidino-2-Benzimidazolyl)Methane Ketone	& -8.47	& 0.61	\\
		DB08656	& 5-amino-2-methyl-N-[(1R)-1-naphthalen-1-ylethyl]benzamide	& -8.47	& 0.61	\\
		DB01976	& Aminoanthracene &	-8.43 &	0.65\\
		DB04716	& 2-(1,1-DIMETHYLETHYL)9-FLUORO-3,6-DIHYDRO-7H-BENZ[H]	& -8.43 &	0.65	\\
		& -IMIDAZ && \\
		& [4,5-F]ISOQUINOLIN-7-ONE &&\\
		DB08066	& N-[3-(1H-BENZIMIDAZOL-2-YL)-1H-PYRAZOL-4-YL]BENZAMIDE	& -8.42 &0.66\\
		DB07457	& 3-[1-(3-AMINOPROPYL)-1H-INDOL-3-YL]-4-(1H-INDOL-3-YL)-1H &	-8.41 &	0.67\\
		& -PYRROLE-2,5-DIONE &&\\
		DB01765	& (5-Oxo-5,6-Dihydro-Indolo[1,2-a]Quinazolin-7-Yl)-Acetic Acid &-8.40 &0.68	\\
		DB08065	& 2-(1H-pyrazol-3-yl)-1H-benzimidazole &-8.40&	0.68\\
		\hline
	\end{tabular}
	\label{Summary2}
\end{table}

\subsection{Binding affinity prediction and ranking of 7012 investigational or off-market  drugs }
Using our validated machine elarning model, we present the binding affinity prediction and ranking of 7012 investigational or off-market  drugs.  We list the top 30 from the investigational or off-market drugs  in Table \ref{Summary2}. A complete list of the predicted values can be found in the Supporting Material.

Compared FDA-approved drugs, investigational or off-market drugs are more promising SARS-CoV-2 inhibitors.   Among them, DEBIO-1347 has a binding affinity of -9.02 kcal/mol (IC$_{50}$: 0.24 $\mu$M). Another top-ranking drug is 3-(1H-BENZIMIDAZOL-2-YL)-1H-INDAZOLE, which has a binding affinity of -9.01 kcal/mol (IC$_{50}$: 0.24 $\mu$M).

\subsection{Pose prediction}
The prediction of binding poses is also another important task in drug discovery. For example, protein-ligand pose and binding affinity predictions are major tasks in D3R Grand Challenges \cite{gaieb2019d3r}. The availability of binding poses enables researchers to understand the molecular mechanism of protein-drug interactions further. In this work, utilizing the MathPose \cite{nguyen2019mathdl} developed by us recently, we predict and analyze the binding poses of our predicted top 3 FDA approved drugs as well as our predicted top 3 investigational or off-market ones.

\section{Discussion}

\subsection{Analysis of predicted top FDA-approved drugs}

\subsubsection{Top 3 FDA-approved drugs}

\begin{figure}[ht!]
    \centering
    \captionsetup{margin=0.9cm}
    \subfigure[Proflavine, -8.37 kcal/mol]{
        \label{fig:DB01123}
        \includegraphics[scale=0.4]{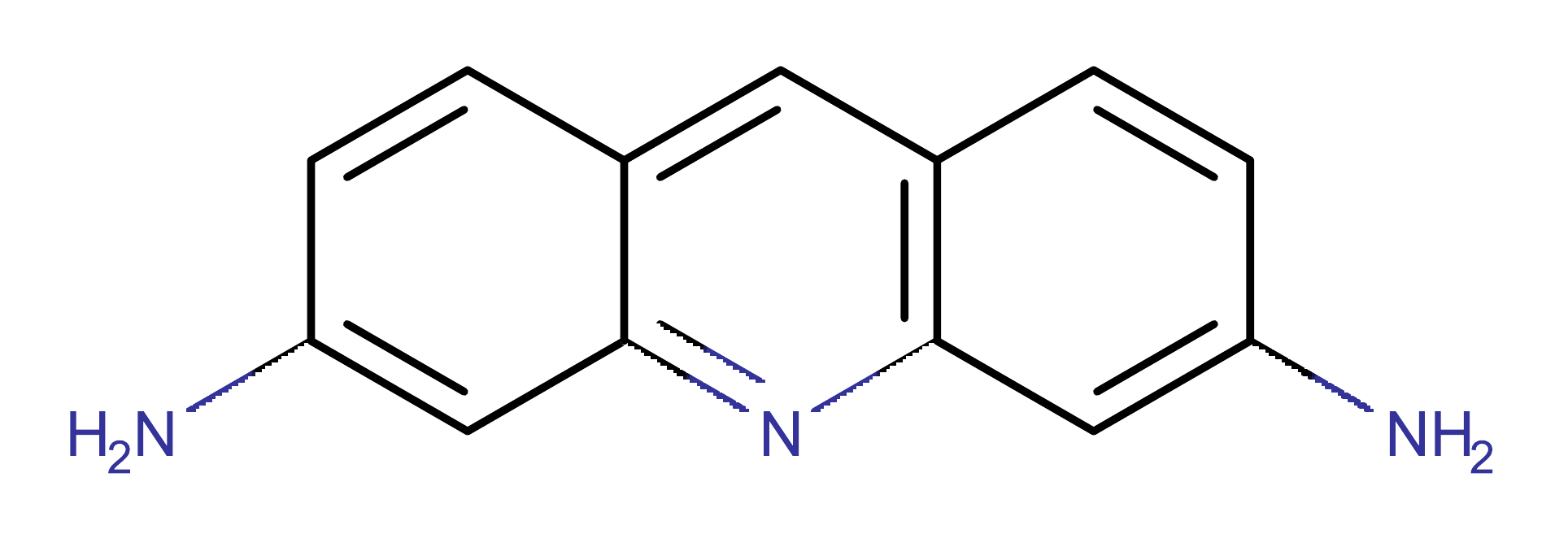}}
        \hspace{0.001in}
    \subfigure[SARS-CoV-2 protease and Proflavine complex]{
        \label{fig:CDB01123}
        \includegraphics[scale = 0.25]{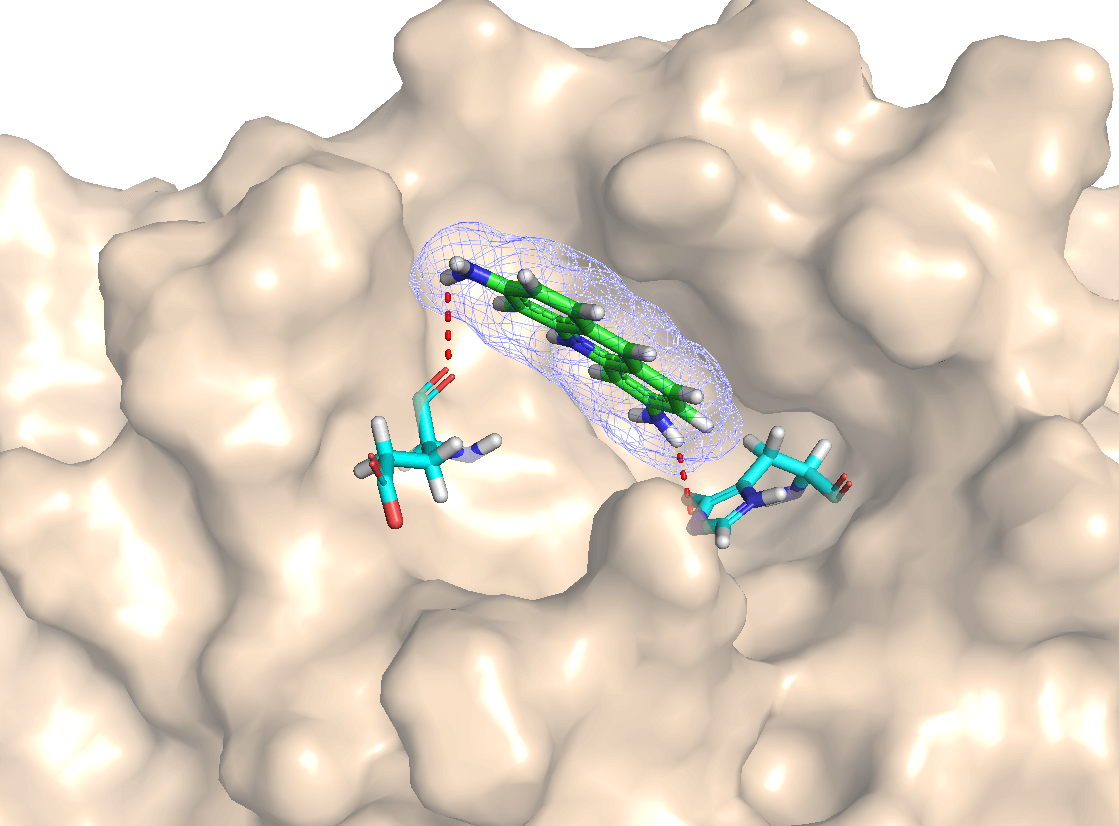}}
        \hspace{0.001in}
    \vfill
    \subfigure[Chloroxine, -8.24 kcal/mol]{
        \label{fig:DB01243}
        \includegraphics[scale=0.93]{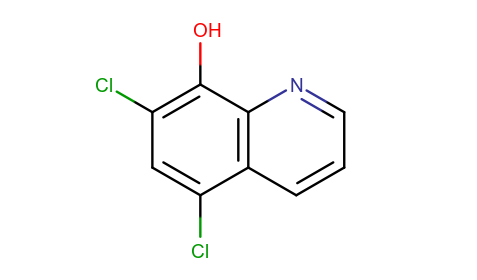}}
        \hspace{0.001in}
    \subfigure[SARS-CoV-2 protease and Chloroxine complex]{
        \label{fig:CDB01243}
        \includegraphics[scale = 0.42]{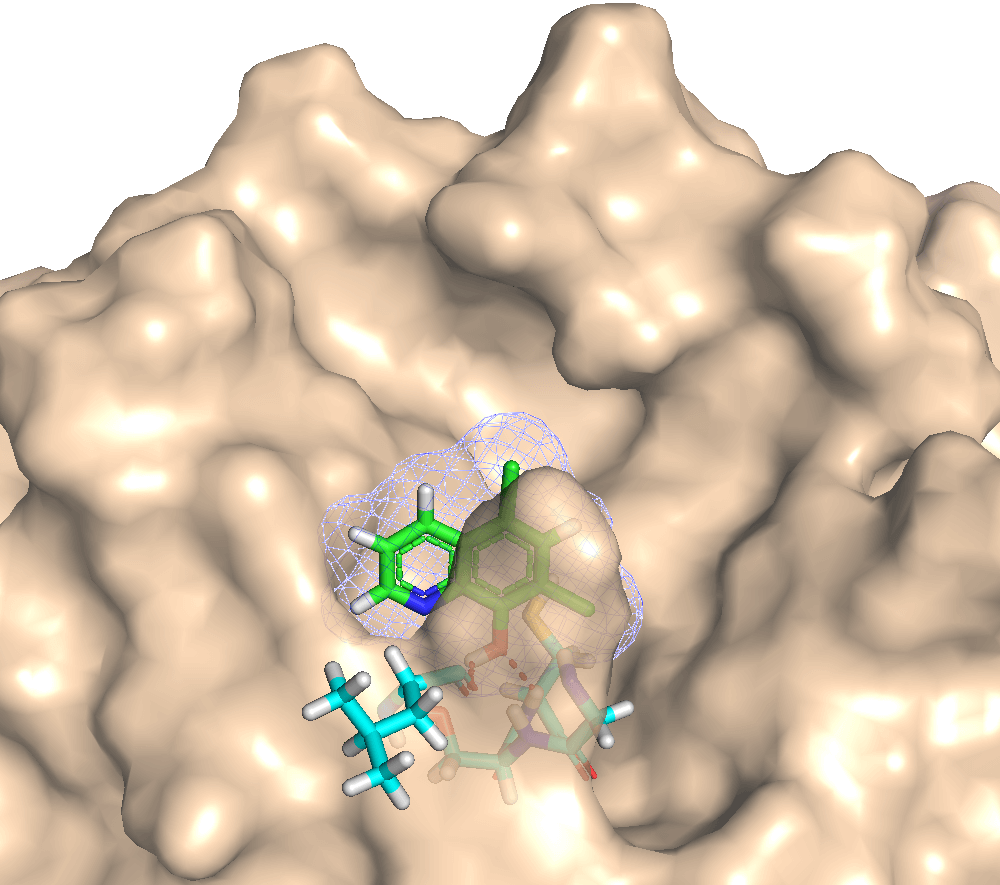}}
        \hspace{0.001in}
    \vfill
    \subfigure[Demexiptiline, -8.14 kcal/mol]{
        \label{fig:DB08998}
        \includegraphics[scale=0.4]{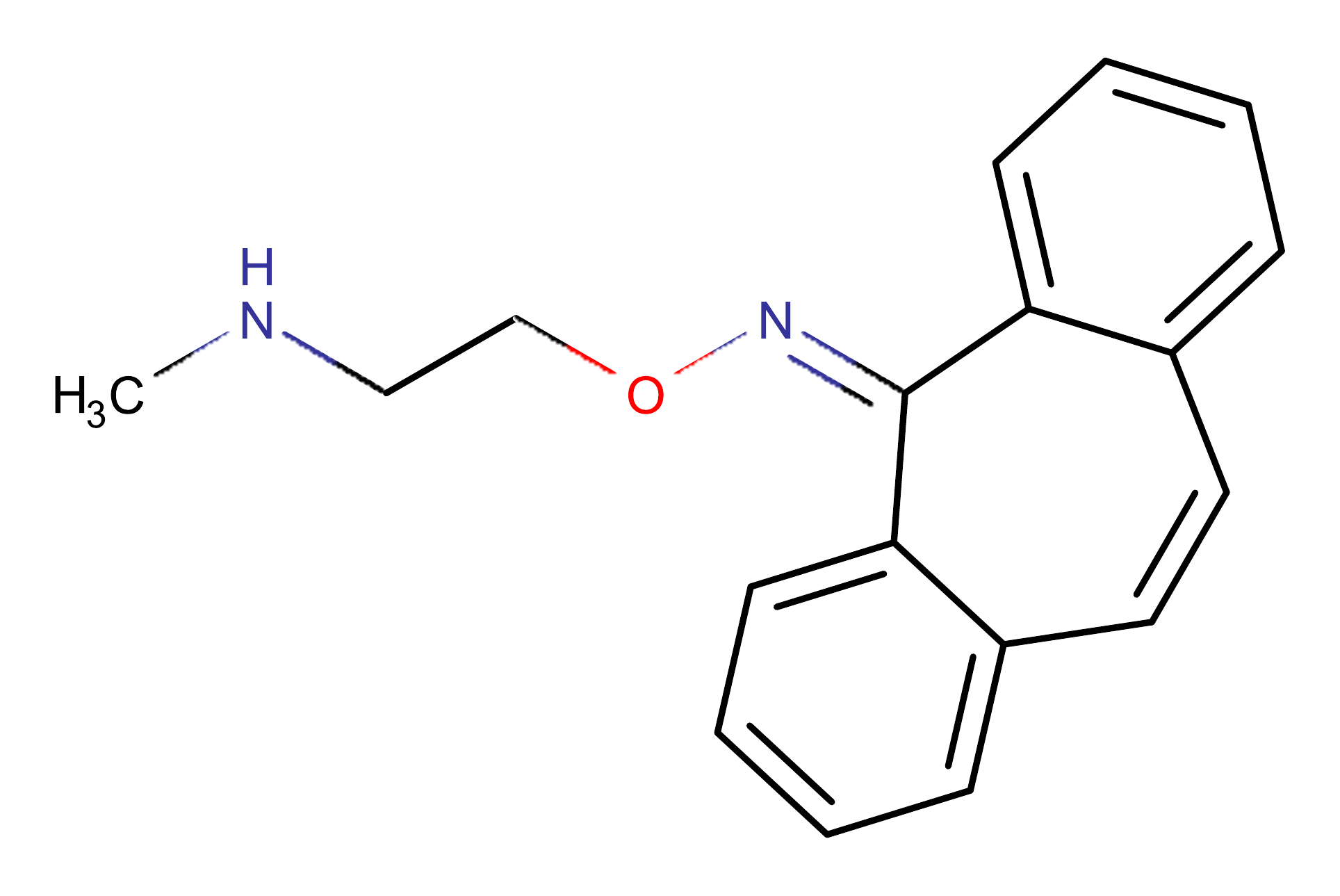}}
        \hspace{0.001in}
    \subfigure[SARS-CoV-2 protease and Demexiptiline complex]{
        \label{fig:CDB08998}
        \includegraphics[scale = 0.29]{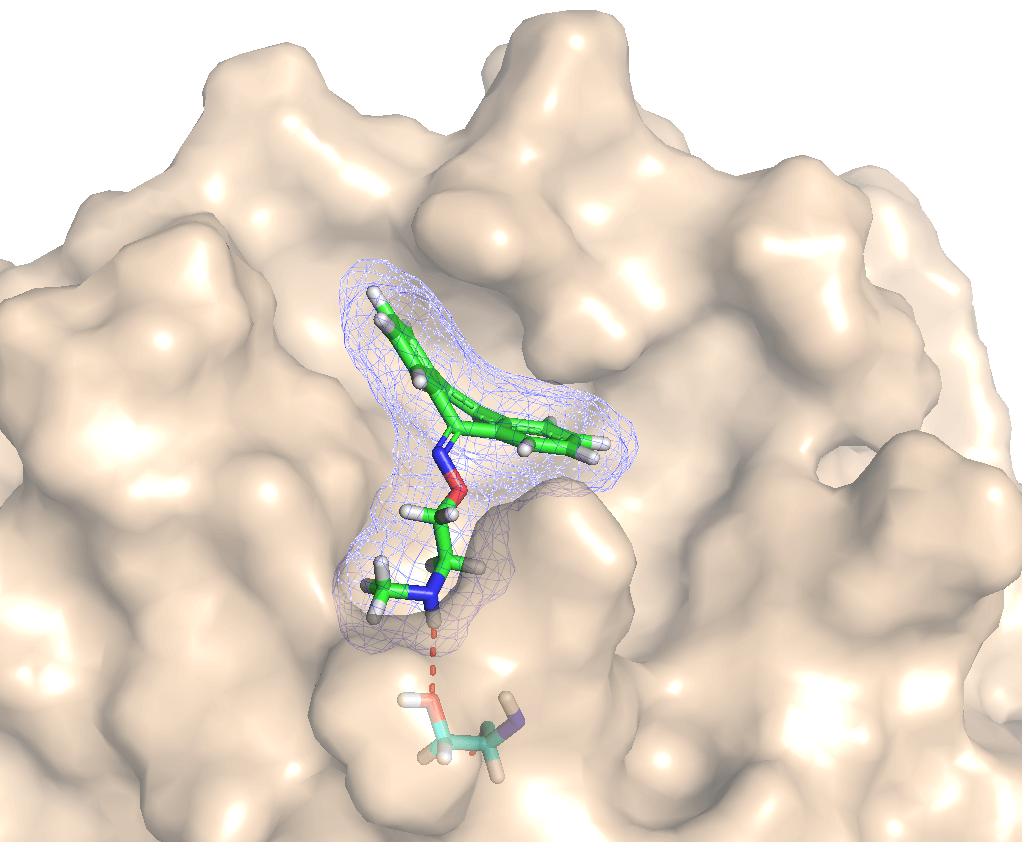}}
        \hspace{0.001in}
    \caption{Proflavine, Chloroxine, Demexiptiline and their complexes with SARS-CoV-2 protease.}
    \label{fig:Top3}
\end{figure}

The first-ranking candidate from the FDA-approved drugs is Proflavine  (see Figure \ref{fig:DB01123}), with a predicted binding affinity to the SARS-CoV-2 main protease of -8.37 kcal/mol. The predicted binding pose using our MathPose \cite{nguyen2019mathdl} is illustrated in Figure \ref{fig:CDB01123}. It reveals that there are two hydrogen bonds formed between the drug and the SARS-CoV-2 main protease. The first one is between one amino of Proflavine and the O atom in the main chain of the residue Glu166 of the protease. The second one is between the other amino of the drug and the five-member ring in the side chain of the residue His41 of the protease.  As a result, the binding affinity is promising.

The predicted second-best drug is Chloroxine (see Figure \ref{fig:DB01243}). Its predicted binding affinity is -8.24 kcal/mol. Between the drug and the protease, there are two hydrogen bonds (see Figure \ref{fig:CDB01243}): One is formed by the H atom of the hydroxy of the drug with the main-chain O atom of the residue Leu141. The other one is between the hydroxy O atom of the drug and the amino in the main chain of Cys145.

The third one, Demexiptiline (see Figure \ref{fig:DB08998}), has a predicted binding affinity of -8.14 kcal/mol. The hydrogen bond between this drug and the protease are formed by the H atom of the amino on the tail of the drug with the side-chain O atom of Ser144. Hydrophobic interactions also play a critical role in the binding.

\subsubsection{The binding affinities of the protease-based drugs}

\begin{table}[ht!]
    \caption{A summary of the predicted binding affinities (BAs) (unit: kcal/mol) and $\text{IC}_{50}$ ($\mu$M) of the existing protease inhibitors. Numbers in parenthesis are predictions from the literature~\cite{beck2020predicting}.}
    \centering
\setlength\tabcolsep{3pt}
\captionsetup{margin=0.9cm}
    \begin{tabular}{l|c|c||l|c|c} \hline
    DrugID &Predicted &$\text{IC}_{50}$&DrugID &Predicted &$\text{IC}_{50}$\\
    &Binding Affinity&& &BA &\\\hline
    Remikiren&-7.42&3.57&Moexipril&-6.55&15.63   \\
    Candoxatril&-7.22&5.05&Trandolapril&-6.54&17.70 \\
    Darunavir&-7.16&5.55&Lopinavir&-6.50&16.92    \\
    Isoflurophate&-7.09&6.28&Spirapril&-6.49 &17.16\\
    Atazanavir&-7.03 (-9.57) &6.96 &Dabigatran etexilate&-6.46&17.96 \\
    Argatroban&-7.02&6.98&Apixaban&-6.44&18.84 \\
    Sitagliptin&-6.93&8.22&Tipranavir&-6.39&20.36 \\
    Fosamprenavir&-6.92&8.26&Lisinopril&-6.35&21.87\\
    Quinapril&-6.91&8.45&Perindopril&-6.34&22.10 \\
    Amprenavir&-6.82&9.83&Cilazapril&-6.31&23.36 \\
    Benazepril&-6.81&10.05&Ritonavir&-6.26 (-8.47)&25.50 \\
    Rivaroxaban&-6.74&11.21&Ximelagatran&-6.24&26.14 \\
    Fosinopril&-6.74&11.28&Vildagliptin&-6.15&30.38 \\
    Telaprevir&-6.73&11.54&Cilastatin&-6.15&30.40 \\
    Captopril&-6.72&11.68&Indinavir&-6.11&32.91 \\
    Ramipril&-6.66&12.84&Saxagliptin&-6.07&35.27 \\
    Enalapril&-6.66&12.93&Nelfinavir&-6.05&36.23\\
    Alogliptin&-6.62&13.90&Boceprevir&-6.00&39.16 \\
    Linagliptin&-6.58&14.73&Simeprevir&-5.77 (-8.29)&58.25 \\
    Saquinavir&-6.56&15.26&Ecabet&-5.71&64.15 \\\hline
    \end{tabular}
    \label{SummaryPI}
\end{table}
It is interesting to analyze the binding affinities of the existing drugs developed as protease inhibitors. Table \ref{SummaryPI} shows their predicted binding affinities. The predicted values by a recent study \cite{beck2020predicting} are given in the parenthesis, it appears that these values are overestimated.  Notably, the current protease inhibitors do not have a substantial effect on the SARS-CoV-2 protease. A possible reason is that SARS-CoV-2 3CL protease is genetically and structurally different from most other known proteases.

\subsubsection{Comparison to experimental data}

\begin{table}[ht!]
\caption{A summary of our predicted binding affinities (BAs) and the corresponding experimental ones of some existing drugs against SARS-CoV-2. All numbers are in kcal/mol unit.}
\centering
\setlength\tabcolsep{3pt}
\captionsetup{margin=0.9cm}
\begin{tabular}{l|c|c||l|c|c} \hline
	DrugID &Experiment &Prediction&DrugID &Experiment &Prediction\\
	\hline
	Remdesivir & -6.74 \cite{jeon2020identification} & -6.29 & Perhexiline & -7.08 \cite{jeon2020identification}& -6.67 \\
	Chloroquine & -7.00 \cite{jeon2020identification} & -6.92 & Loperamide  & -6.86 \cite{jeon2020identification}& -6.98 \\
	Lopinavir & -6.87 \cite{jeon2020identification} & -6.51 & Mefloquine & -7.31 \cite{jeon2020identification}& -6.89 \\
	Niclosamide & -8.93 \cite{jeon2020identification} & -7.66 & Amodiaquine & -7.21 \cite{jeon2020identification} & -6.93 \\
	Proscillaridin & -7.75 \cite{jeon2020identification} & -6.50 & Phenazopyridine & -6.21 \cite{jeon2020identification} & -7.51\\
	Penfluridol & -7.23 \cite{jeon2020identification} & -6.54 & Clomiphene & -7.19 \cite{jeon2020identification}& -7.12\\
	Toremifene & -7.42 \cite{jeon2020identification} & -7.20 & Digoxin & -9.16 \cite{jeon2020identification} & -7.00\\
	Hexachlorophene & -8.24 \cite{jeon2020identification} & -7.37 & Thioridazine & -7.05 \cite{jeon2020identification} & -6.96\\
	Salinomycin & -9.02 \cite{jeon2020identification} & -7.00 & Pyronaridine & -6.13 \cite{jeon2020identification} & -6.68\\
	Ciclesonide & -7.31 \cite{jeon2020identification} & -7.04 & Ceritinib  & -7.56 \cite{jeon2020identification}& -6.77\\
	Osimertinib & -7.48 \cite{jeon2020identification} & -6.62 & Lusutrombopag & -7.39 \cite{jeon2020identification} & -6.78\\
	Gilteritinib & -7.05 \cite{jeon2020identification} & -5.57 & Berbamine & -6.96 \cite{jeon2020identification} & -6.87\\
	Ivacaftor & -7.07 \cite{jeon2020identification} & -6.74 & Mequitazine & -7.00 \cite{jeon2020identification} & -6.41\\
	Dronedarone & -7.37 \cite{jeon2020identification} & -6.19 & Eltrombopag & -6.93 \cite{jeon2020identification} & -6.17\\
	Fluphenazine & -7.08 \cite{weston2020fda} & -6.29 & Benztropine & -6.63 \cite{weston2020fda} & -6.94\\
	Chlorpromazine & -7.50 \cite{weston2020fda} & -7.00& Terconazole & -6.71 \cite{weston2020fda} & -7.18 \\
	Simeprevir & -6.67 \cite{ma2020boceprevir} & -5.77& Boceprevir & -7.34 \cite{ma2020boceprevir} & -6.00 \\
	Narlaprevir & -7.14 \cite{ma2020boceprevir} & -6.38& &  &  \\

	\hline
\end{tabular}
\label{pred_vs_exp}
\end{table}

In this section, we are interested in comparing our predicted binding affinities to the corresponding experimental ones of some existing drugs outside our training set. Table \ref{pred_vs_exp} lists our predictions along with the experimental values of these drugs. These experimental data are extracted from the recent literature \cite{ma2020boceprevir,jeon2020identification,weston2020fda}. The RMSE  of experimental values and predicted ones is 0.87 kcal/mol, showing a good agreement.
It is to point out that all these data were obtained from cell-culture experiments, leading to discrepancies when comparing these experimental values to our results only tailoring to the inhibition of the SARS-CoV-2 3CL protease. For example, the target of Remdesivir is the RNA-dependent RNA polymerase rather than the 3CL protease.

\begin{figure}[ht!]
	\centering
	\captionsetup{margin=0.9cm}
	\subfigure[DEBIO-1347, -9.02 kcal/mol]{
		\label{fig:DB12903}
		\includegraphics[scale=0.38]{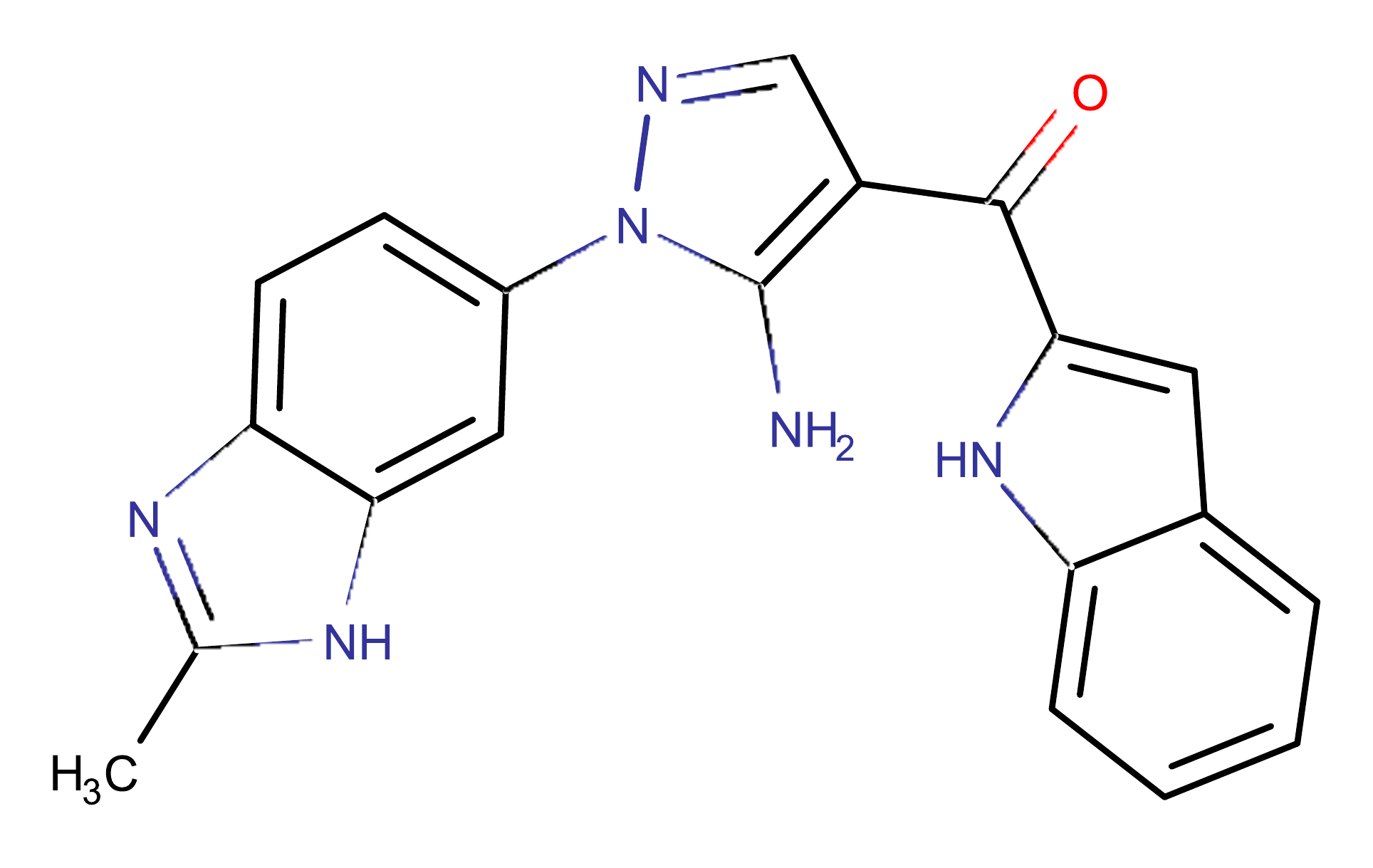}}
	\hspace{0.001in}
	\subfigure[SARS-CoV-2 protease and DEBIO-1347 complex]{
		\label{fig:CDB12903}
		\includegraphics[scale = 0.38]{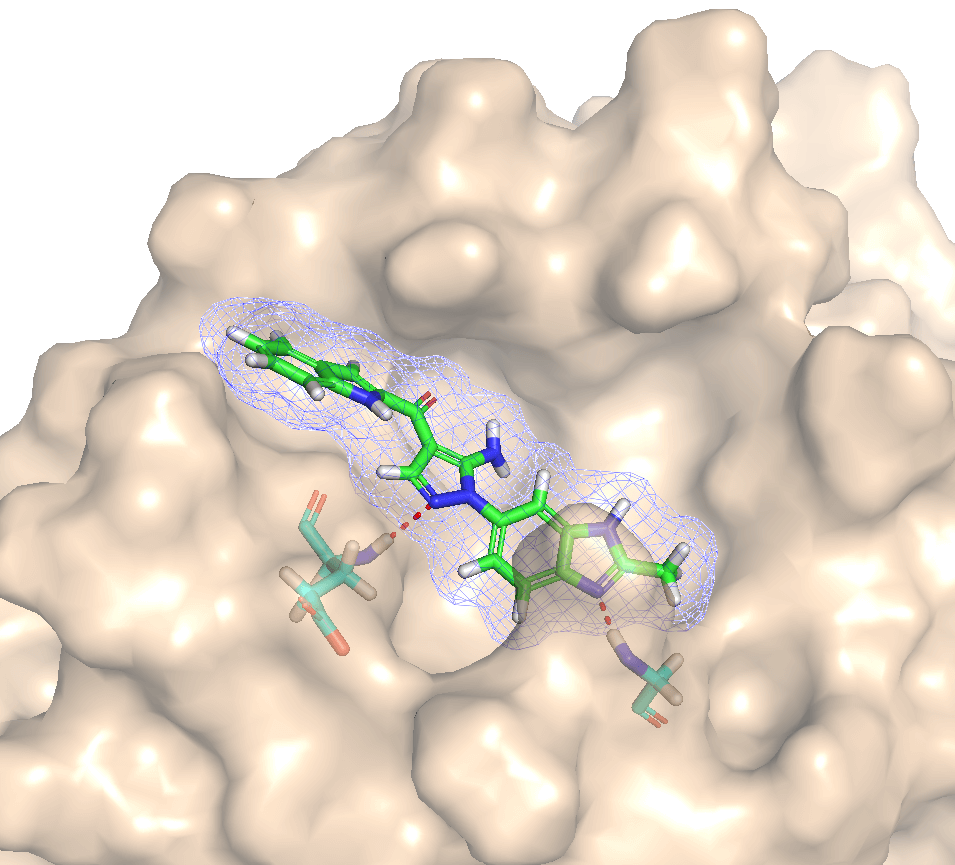}}
	\hspace{0.001in}
	\vfill
	\subfigure[3-(1H-BENZIMIDAZOL-2-YL)-1H-INDAZOLE, \newline -9.01 kcal/mol]{
		\label{fig:DB07959}
		\includegraphics[scale=0.38]{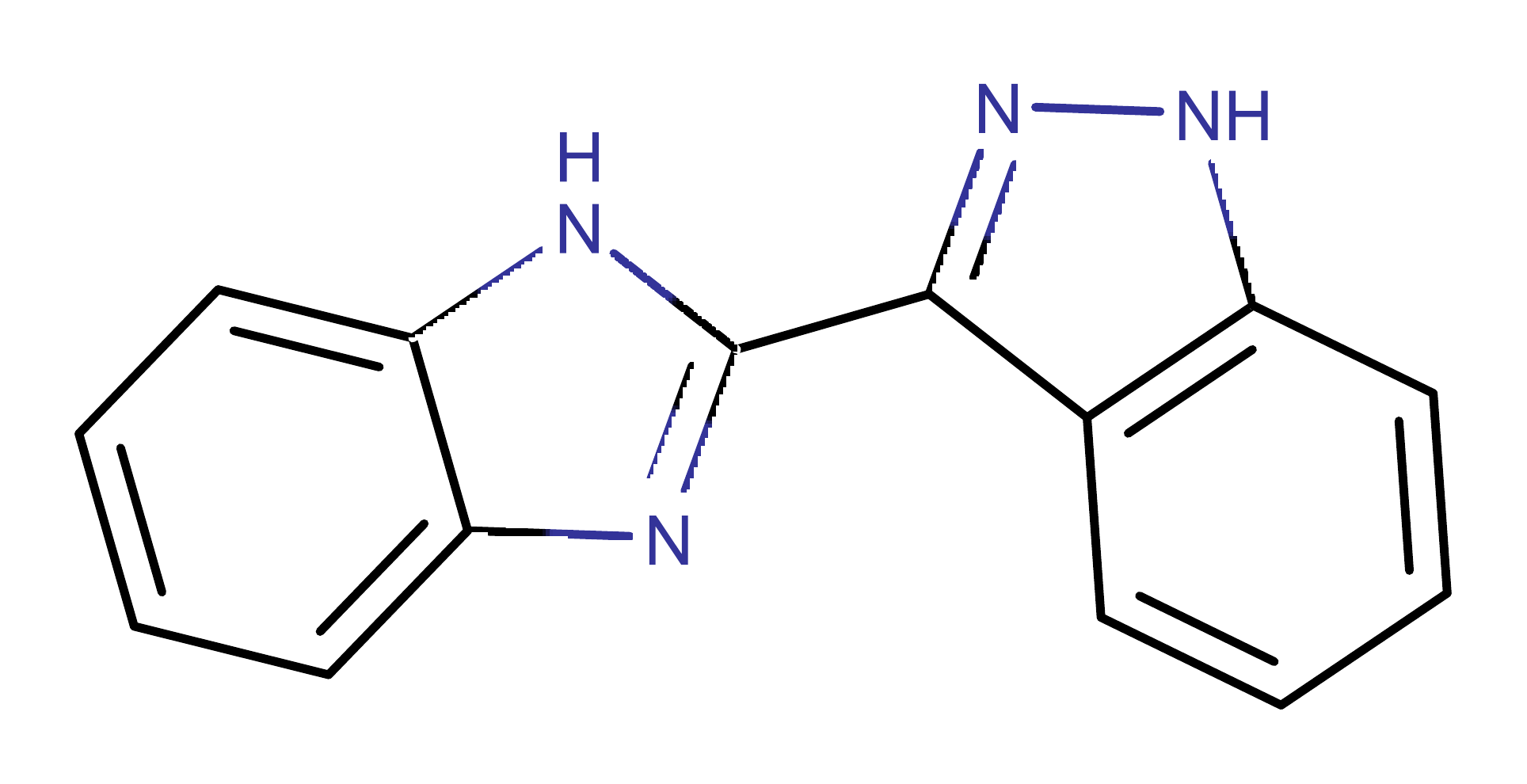}}
	\hspace{0.001in}
	\subfigure[SARS-CoV-2 protease and 3-(1H-BENZIMIDAZOL-2-YL)-1H-INDAZOLE complex]{
		\label{fig:CDB07959}
		\includegraphics[scale = 0.38]{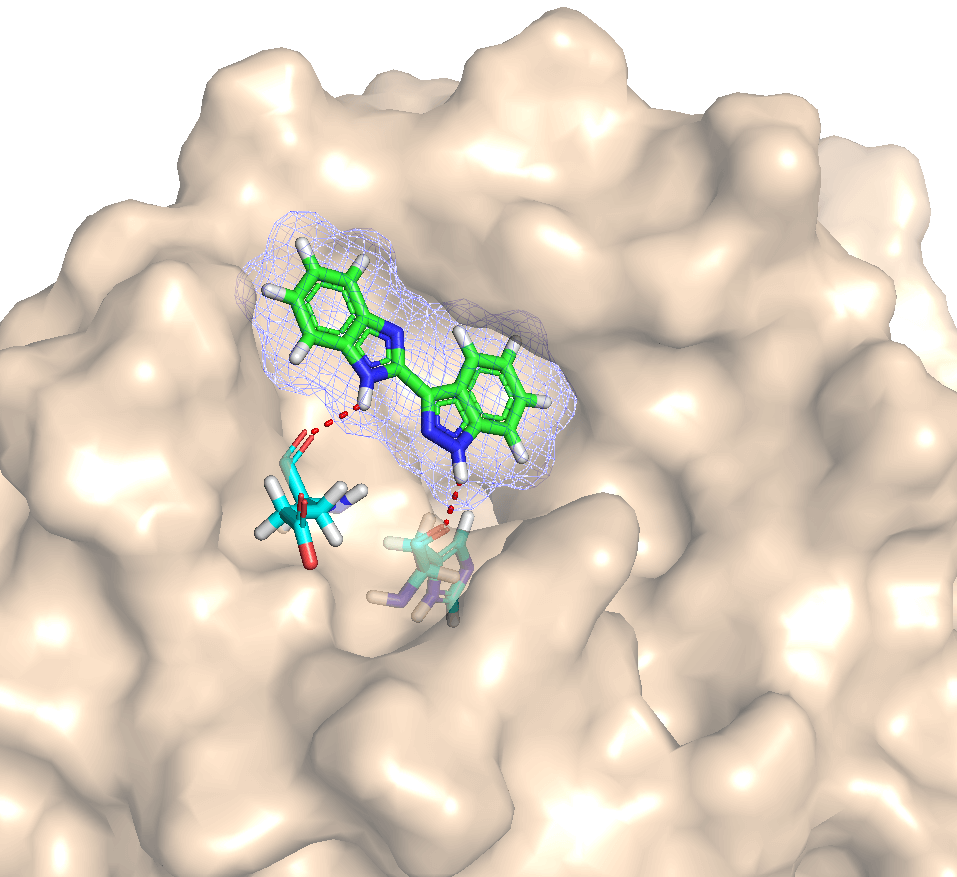}}
	\hspace{0.001in}
	\vfill
	\subfigure[9H-CARBAZOLE, -8.96 kcal/mol]{
		\label{fig:DB07301}
		\includegraphics[scale=0.38]{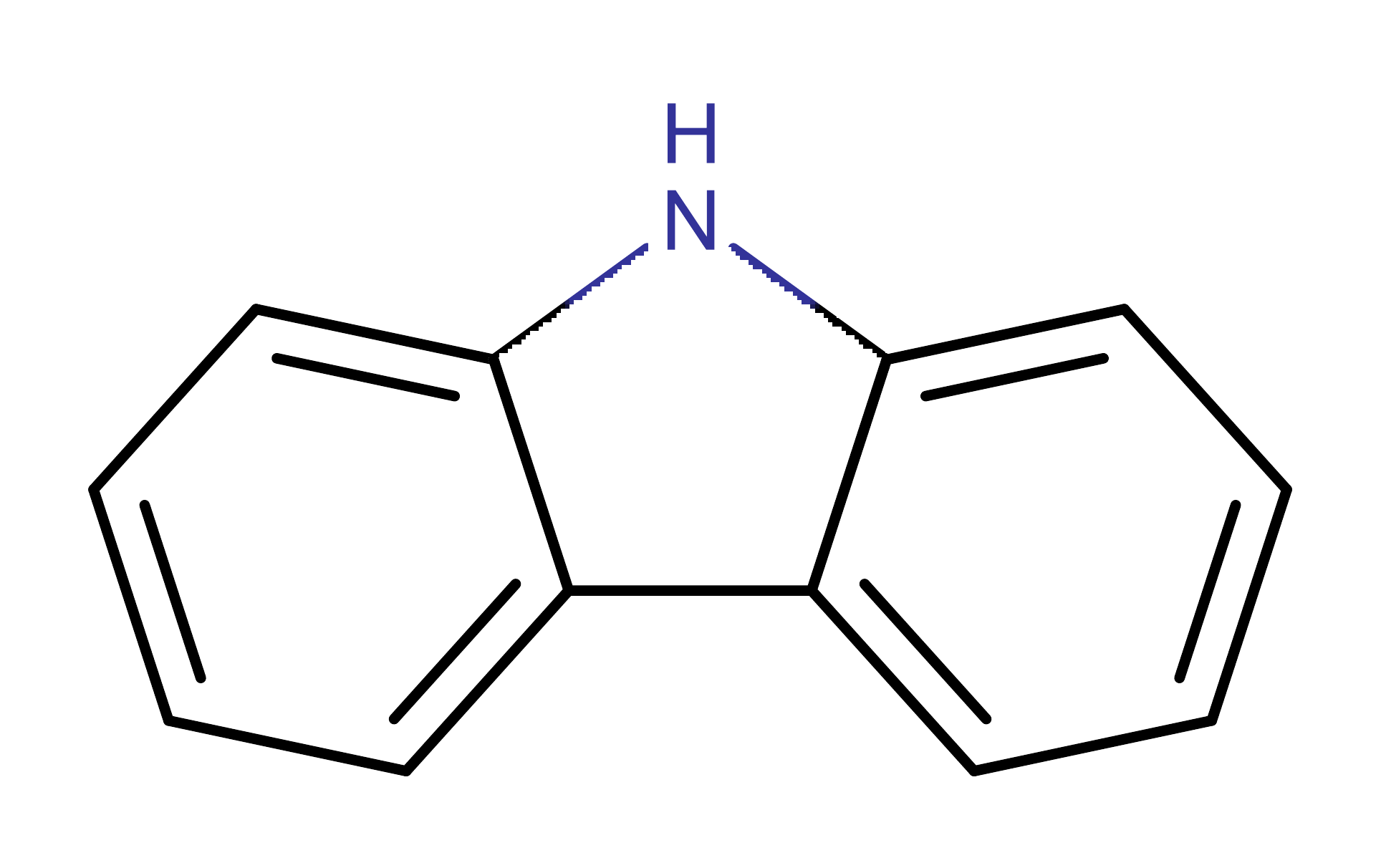}}
	\hspace{0.001in}
	\subfigure[SARS-CoV-2 protease and 9H-CARBAZOLE complex]{
		\label{fig:CDB07301}
		\includegraphics[scale = 0.38]{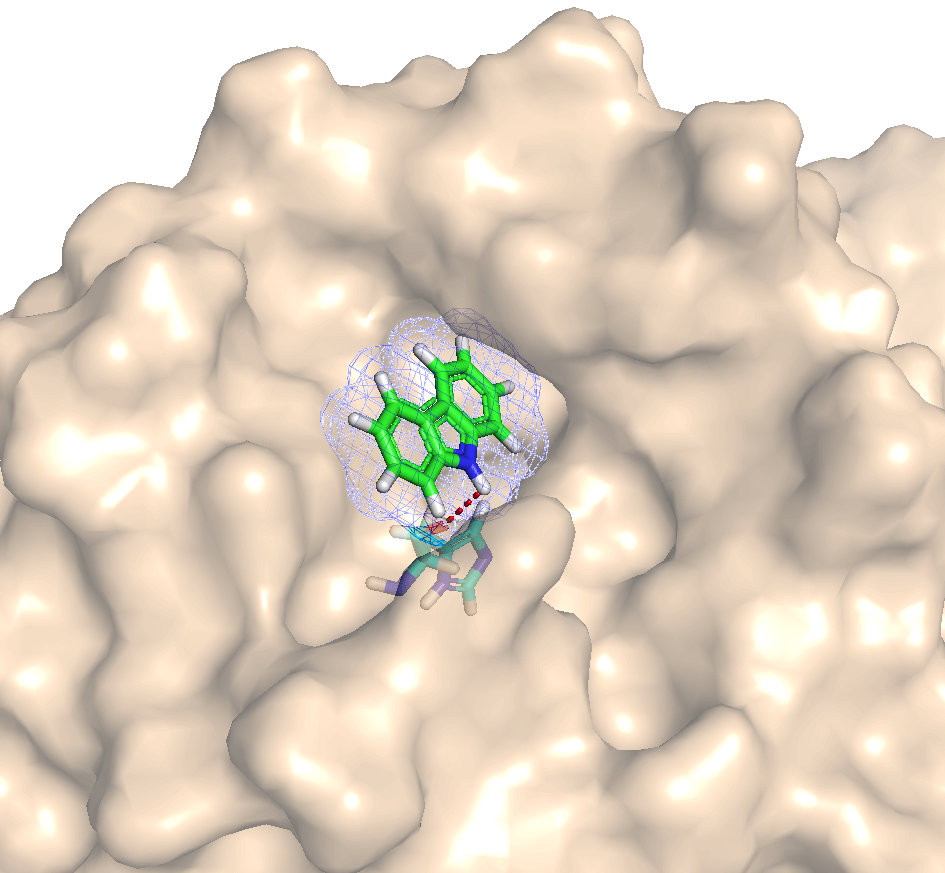}}
	\hspace{0.001in}
	\caption{DEBIO-1347, 3-(1H-BENZIMIDAZOL-2-YL)-1H-INDAZOLE, 9H-CARBAZOLE and their complexes with SARS-CoV-2 protease.}
	\label{fig:Top3-2}
\end{figure}

\subsection{Analysis of  top 3 investigational or off-market drugs}

Among the investigational or off-market drugs, the top-ranking candidate is DEBIO-1347  (see Figure \ref{fig:DB12903}). Its binding affinity with the SARS-CoV-2 protease is predicted to be -9.02 kcal/mol. The MathPose predicted pose is illustrated in Figure \ref{fig:CDB12903}. It indicates a hydrogen bond network formed between the drug and the protease leads to the moderately high binding affinity. This network consists of two hydrogen bonds: the first hydrogen bond is between one N atom in the Pyrazole of the drug and the main-chain amino of the residue Glu166 of the protease; the second one is between one N atom in the 1H-1,3-benzodiazole of the drug and the main-chain amino of the residue Gly143 of the protease.

The second-best investigational drug is 3-(1H-BENZIMIDAZOL-2-YL)-1H-INDAZOLE (Figure \ref{fig:DB07959}) with a  predicted binding affinity of -9.01 kcal/mol. Figure \ref{fig:CDB07959} reveals that the drug forms two hydrogen bonds with the protease. One is between one N atom in the 1H-1,3-benzodiazole of the drug and the main-chain O atom of the residue Glu166 of the protease. The other is between one N atom in the 1H-indazole of the drug and the main-chain O atom of the residue His164 of the protease.

The third one, 9H-CARBAZOLE (see Figure \ref{fig:DB07301}), also has a promising predicted affinity of -8.96 kcal/mol. One can see from Figure \ref{fig:CDB07301}, a strong hydrogen bond is formed between the N atom of the drug and the main-chain O atom of the residue His164 of the protease. The hydrophobic interactions play an essential role in the binding as well.

\subsection{Analysis of top SARS-CoV/SARS-CoV-2 3CL-protease inhibitors }

Note that, in our training set collected from the existing experimental data, 21 samples have binding affinity values lower than -9 kcal/mol. Table \ref{Summary-trainset} provides a list top 30  SARS-CoV/SARS-CoV-2 3CL-protease inhibitors with their experimental binding affinities and estimated  druggable properties. 
 Moreover, 4 of these 21 samples have 3D experimental structures available. Although these inhibitors are not on the market yet, they serve as good starting points for the design of anti-SARS-CoV-2 drugs.
A full list of our training compounds is given in the Supporting Material.

\subsubsection{Binding interaction analysis of the top 3 experimental structures}

\begin{figure}[ht!]
	\centering
	\captionsetup{margin=0.9cm}
	\subfigure[2zu4 inhibitor, -10.12 kcal/mol]{
		\label{fig:lig-2zu4}
		\includegraphics[scale=0.4]{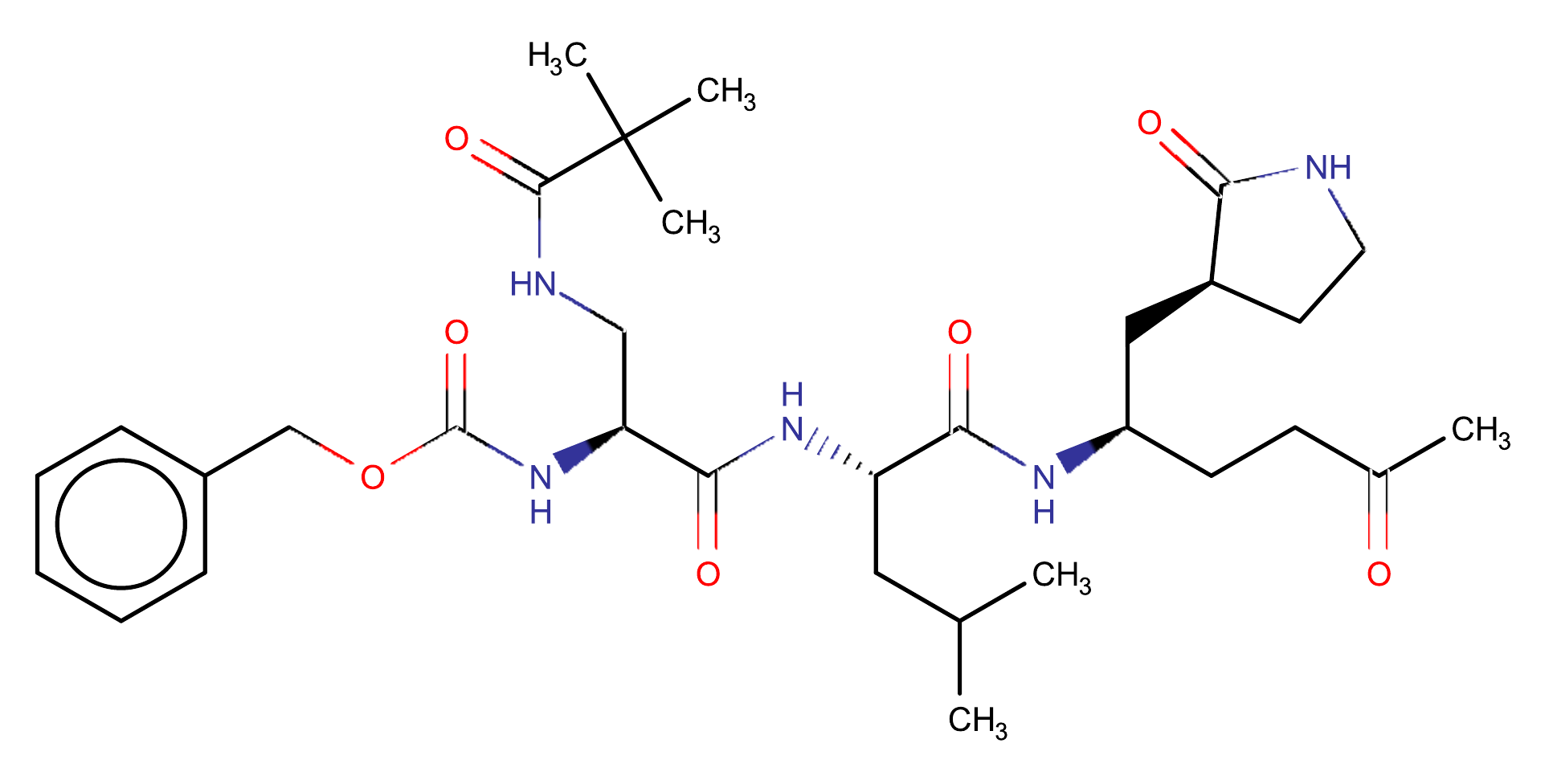}}
	\hspace{0.001in}
	\subfigure[2zu4 complex]{
		\label{fig:2zu4}
		\includegraphics[scale = 0.40]{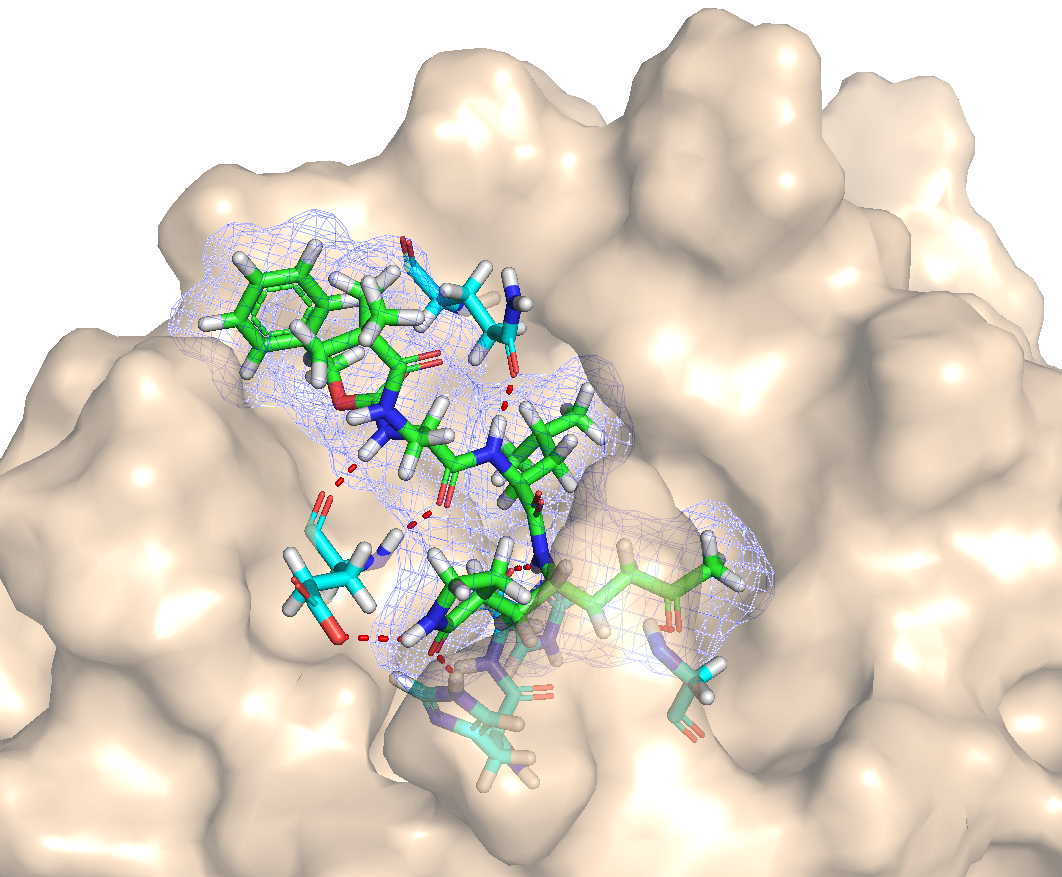}}
	\hspace{0.001in}
	\vfill
	\subfigure[3avz inhibitor, -9.80 kcal/mol]{
		\label{fig:lig-3avz}
		\includegraphics[scale=0.40]{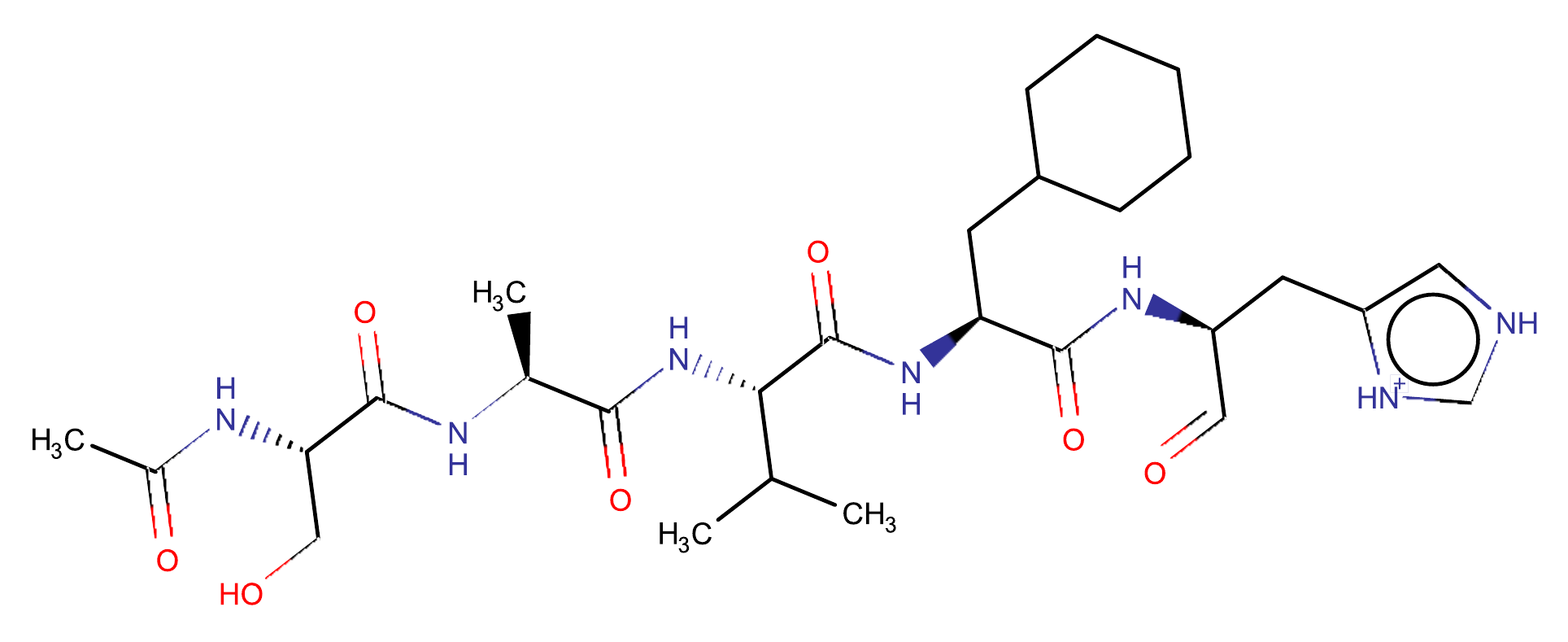}}
	\hspace{0.001in}
	\subfigure[3avz complex]{
		\label{fig:3avz}
		\includegraphics[scale = 0.45]{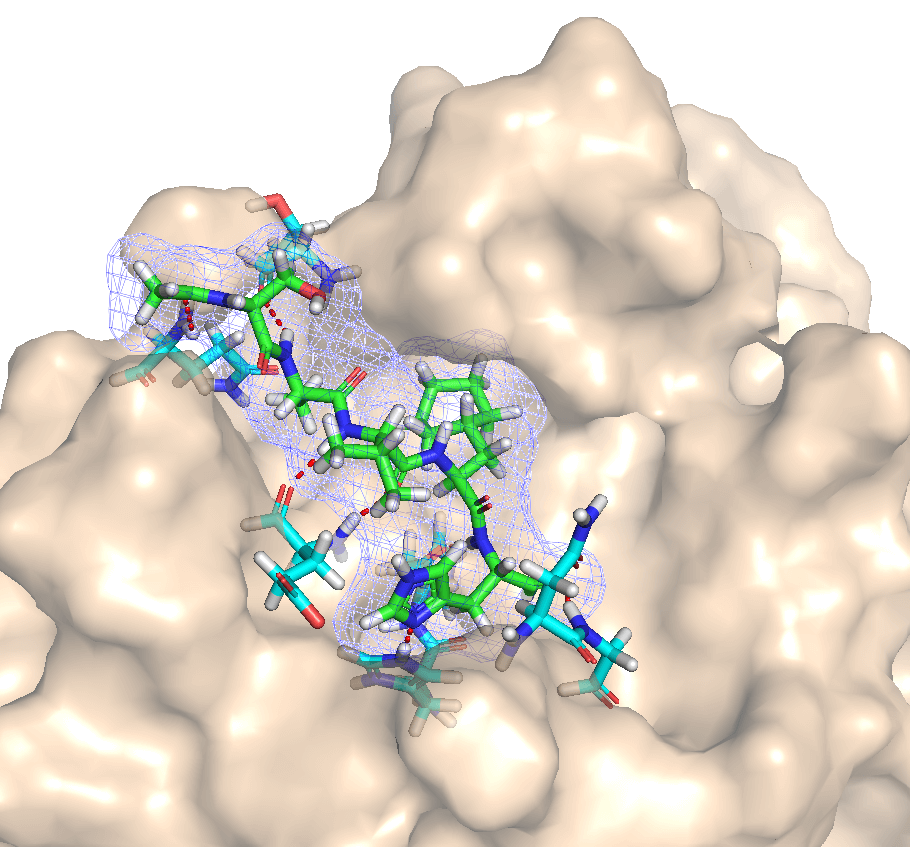}}
	\hspace{0.001in}
	\vfill
	\subfigure[2zu5 inhibitor, -9.56 kcal/mol]{
		\label{fig:lig-2zu5}
		\includegraphics[scale=0.4]{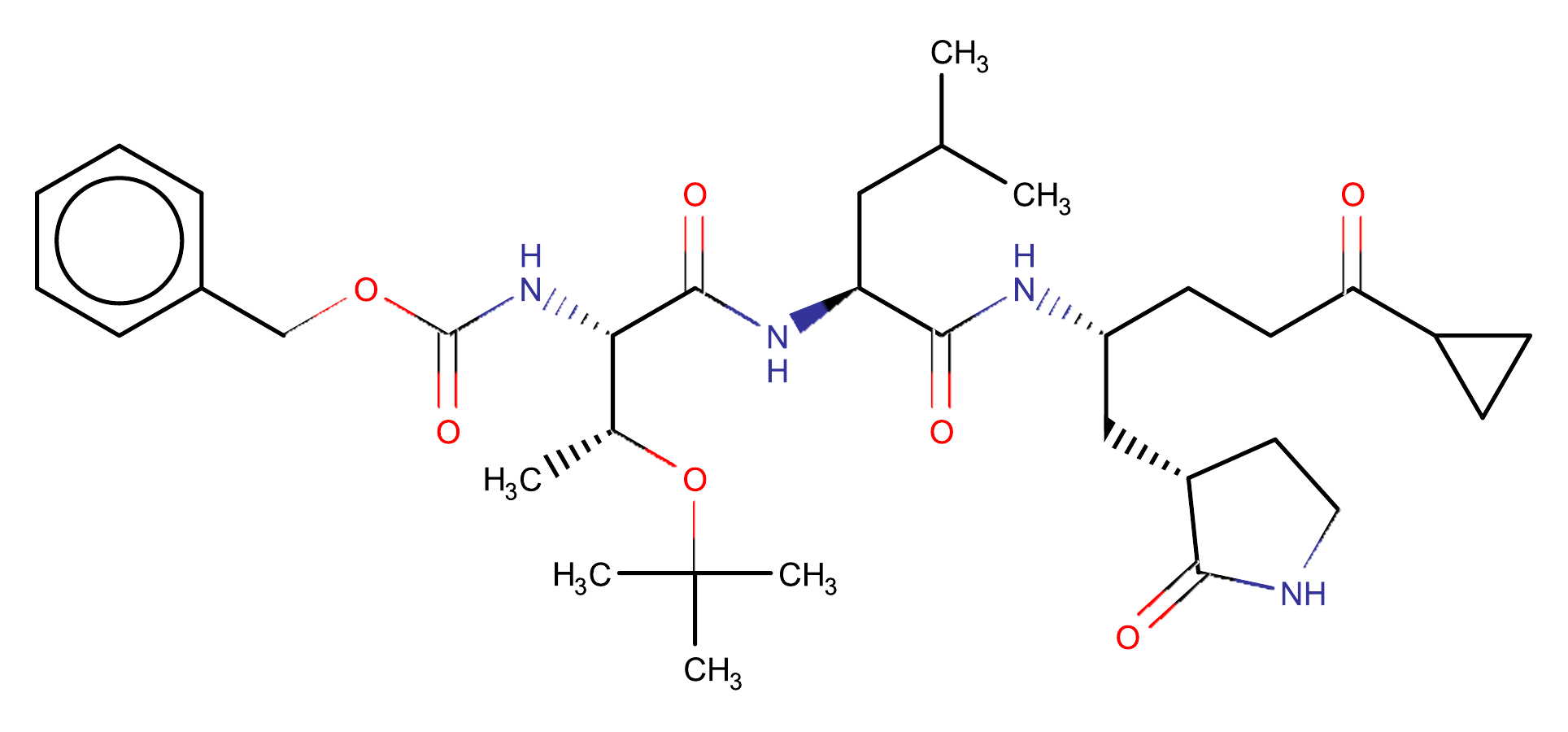}}
	\hspace{0.001in}
	\subfigure[2zu5 complex]{
		\label{fig:2zu5}
		\includegraphics[scale = 0.40]{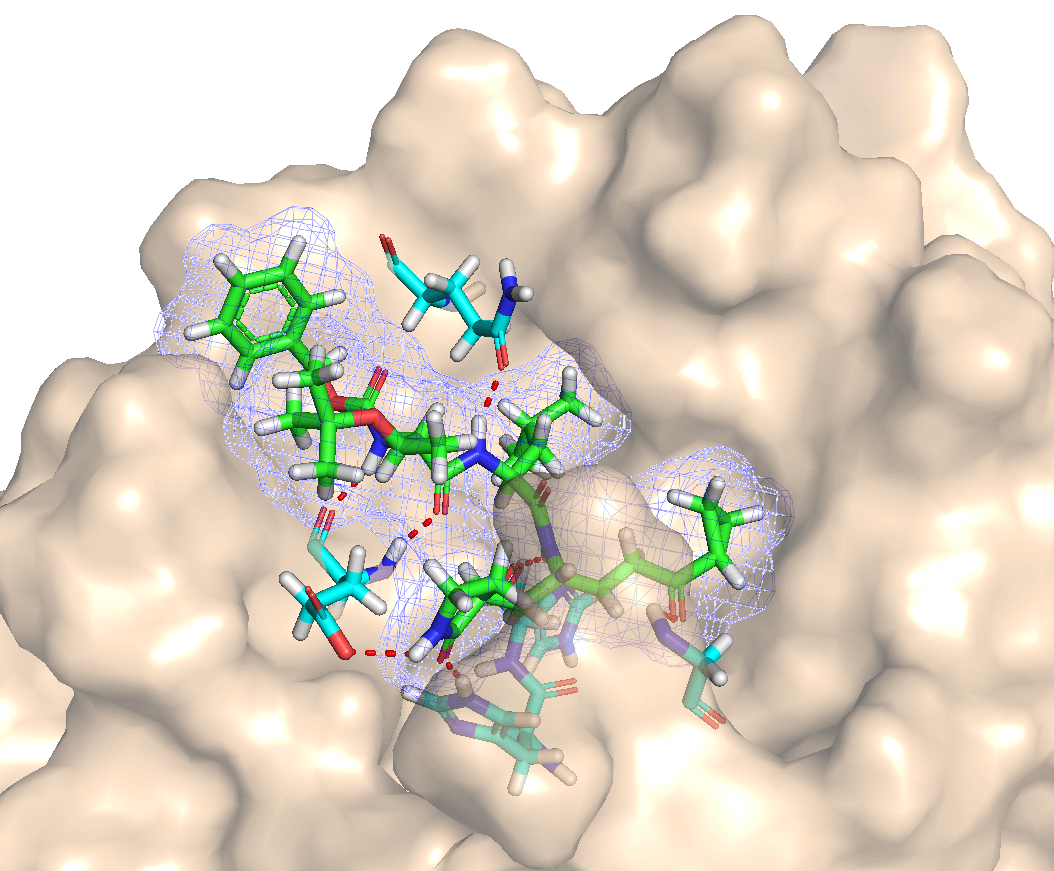}}
	\hspace{0.001in}
	\caption{The inhibitors and complexes from the top 3 PDBbind structures, 2zu4, 3avz, and 2zu5.}
	\label{fig:Top3-3}
\end{figure}

 Among the SARS-CoV/SARS-CoV-2 3CL-protease complexes with their 3D experimental structures available, the one with the PDB ID 2zu4 \cite{lee2009structural} is the most potent one with a binding affinity over -10 kcal/mol. This high binding affinity is due to a strong hydrogen bond network between the inhibitor and the protease, which consists of as many as 7 hydrogen bonds. These 7 hydrogen bonds are formed by the inhibitor with protease residues Gln189, Gly143, His163, His164, and Glu166 of the protease.

The second-best 3D-experimental structure is the one with the PDB ID 3avz \cite{akaji2011structure} and its binding affinity is -9.80 kcal/mol. A hydrogen bond network, including 7 bonds, plays an essential role in this strong binding. This network is between the inhibitor and protease residues Gln192, Thr190, His164, His163, Glu166, and Gly143.

The PDB ID of the third one is 2zu5 \cite{lee2009structural} with a binding affinity of -9.56 kcal/mol. A strong hydrogen bond network with 7 bonds can also be found in the structure. The protease residues in the network are Glu166, Phe148, His163, His164, Gly143, and Gln189.

Since His163, His164, and  Glu166 emerge in the hydrogen bond networks of all these three structures, it suggests that these three residues are critical to inhibitor binding.

\subsubsection{The analysis of log P, log S, and synthesizability of  3 top SARS-CoV/SARS-CoV-2 3CL-protease inhibitors inhibitors }

\begin{table}[ht!]
	\caption{A summary of top 30 SARS-CoV-2 main protease inhibitors in the training set with experimental binding affinities (unit: kcal/mol), $\text{IC}_{50}$ ($\mu$M), as well as calculated synthesizability, log P and log S.}
	\centering
	\setlength\tabcolsep{10pt}
	\captionsetup{margin=0.9cm}
	\begin{tabular}{c|c|c|c|c|c} \hline
		ID & Binding Affinity & $\text{IC}_{50}$ & synthesizability & log P & log S \\ \hline
		CHEMBL497141 & 	-11.08 & 0.01 &	2.4	& 2.18 & -3.65 \\
		PDB ID 2zu4 &	-10.12	& 0.04 & 4.04 & 2.35 & -3.53 \\
		CHEMBL222234 & -9.95 & 0.05	& 2.26 & 2.66 & -3.59 \\
		CHEMBL2442057 &	-9.94 & 0.05 & 2.26 & 5.39 & -6.22 \\
		CHEMBL213054 & -9.92 & 0.05	& 4.2 & 3.15 & -3.81 \\
		CHEMBL212080 & -9.87 & 0.06	& 4.25 & 3.15 & -3.76 \\
		CHEMBL222840 & -9.85 & 0.06	& 2.23 & 2.55 & -3.37 \\
		CHEMBL398437 & -9.85 & 0.06	& 2.29 & 4.12 & -5.39 \\
		CHEMBL222769 & -9.82 & 0.06	& 2.16 & 4.87 & -5.73 \\
		PDB ID 3avz &	-9.80 & 0.07	& 4.65 & -1.35 & -2.33 \\
		CHEMBL225515 &	-9.80 & 0.07	& 2.22	& 3.44	& -4.28 \\
		CHEMBL1929019 &	-9.80 &	0.07 & 4.23	& -0.77	& -2.41 \\
		CHEMBL222893 &	-9.57 & 0.10 &	2.21 &	4.17 &	-5.01 \\
		PDB ID 2zu5	& -9.56	& 0.10	& 4.27	& 3.79	& -4.39 \\
		PDB ID 3atw	 & -9.55 & 0.10	& 4.63	& -0.46	& -2.47 \\
		CHEMBL334399 &	-9.50 & 0.11 & 2.20 & 3.06	& -4.17 \\
		CHEMBL253905	& -9.43	& 0.12	& 2.43	& 4.78 & -5.45 \\
		CHEMBL403932	&	-9.42 & 0.12 &	1.94 &	4.11 & -4.97 \\
		CHEMBL254103	&	-9.25 &	0.16 &	2.10 &	2.35 &	-3.34 \\
		CHEMBL426898	&	-9.23 &	0.17 &	2.17 &	3.70 &	-4.72 \\
		CHEMBL402379	&	-9.11 & 0.21 & 2.34	& 4.12 & -5.36 \\
		CHEMBL222628	&	-8.96 &	0.27 &	2.66 &	2.41 &	-3.52 \\
		CHEMBL1929020	&	-8.96 &	0.27 &	3.97 &	0.76 &	-2.58 \\
		CHEMBL212218	&	-8.89 &	0.30 &	2.71 &	4.97 &	-5.16 \\
		PDB ID 2gz7	&	-8.89	& 0.30	& 2.71	& 4.97	& -5.16 \\
		PMC2597651\_1 & 	-8.88 &	0.31 &	2.45 &	4.49 &	-5.15 \\
		CHEMBL2402686	&	-8.84 &	0.33 &	3.69 &	3.14 &	-4.15 \\
		CHEMBL252662	&	-8.83 &	0.33 &	2.11 &	2.86 &	-4.06 \\
		CHEMBL222735	&	-8.82 &	0.34 &	1.91 &	2.96 &	-3.84 \\
		CHEMBL1929022	&	-8.82 &	0.34 &	4.01 &	-0.27 &	-2.42 \\
		\hline
	\end{tabular}
	\label{Summary-trainset}
\end{table}

The partition coefficient (log P), aqueous solubility (log S), and synthesizability are also critical medical chemical properties for deciding whether a compound can be a drug or not. Notably, synthesizability is always in terms of synthetic accessibility score (SAS), for which 1 means the easiest, 10 means the hardest. Here, we first calculate the log Ps, log Ss, and SASs of the 1553 FDA-approved drugs (See the Supporting Material), then investigate whether the three properties of the inhibitors in the top 3 experimental structures are in the preferred ranges of that of the FDA-approved drugs.

According to the log P distribution of the FDA-approved drugs in the Supporting Material, the log P interval with a large population of the FDA-approved drugs is between -0.14 and 4.96. The log P values of the top 3 inhibitors are 2.35, -1.35, and -0.46, respectively.

The log S distribution reveals that the preferred range of log S is between -5.12 and 1.76. The log S values of the top 3 inhibitors are -3.53, -2.33, and -4.39.

In the SAS distribution, most of the FDA-approved drugs have SASs between 1.84 and 3.94. The SAS values of the top 3 inhibitors are 4.04, 4.65, and 4.27.

So in summary, for the inhibitor in the first ranking PDB structure 2zu4, its log P and log S are quite good for a drug. The SAS is a little higher, but it is still not so hard to synthesize: 344 of the 1553 FDA-approved drug have larger SASs than this inhibitor, even 56 of them have SASs over 6.

Similarly, for the 3avz and 2zu5 inhibitors, their log Ss are very promising. The log Ps and SASs are some out of the preferred ranges, but still, many FDA-approved drugs have worse log Ps and SASs than theirs. As a result, these top 3 inhibitors, especially the first one, could be good starting points for developing anti-SARS-CoV2 drugs. Obviously, their toxicity will be a major concern for any further development.

\section{Material and methods}

\subsection{Sequence identity and 3D structure similarity analysis}

 The sequence identity is defined as the percentage of characters that match exactly between two different sequences. Calculated by SWISS-MODEL\cite{bienert2017swiss}, the sequence identities between the SARS-CoV-2 3CL protease and that of SARS-CoV, MERS-CoV, HKU-1, OC43, HCoVNL63, 229E, and HIV are 96.1\%, 52.0\%, 49.0\%, 48.4\%, 45.2\%, 41.9\%, and 23.7\%, respectively.

\begin{figure}[ht!]
	\centering
	\includegraphics[scale=0.5]{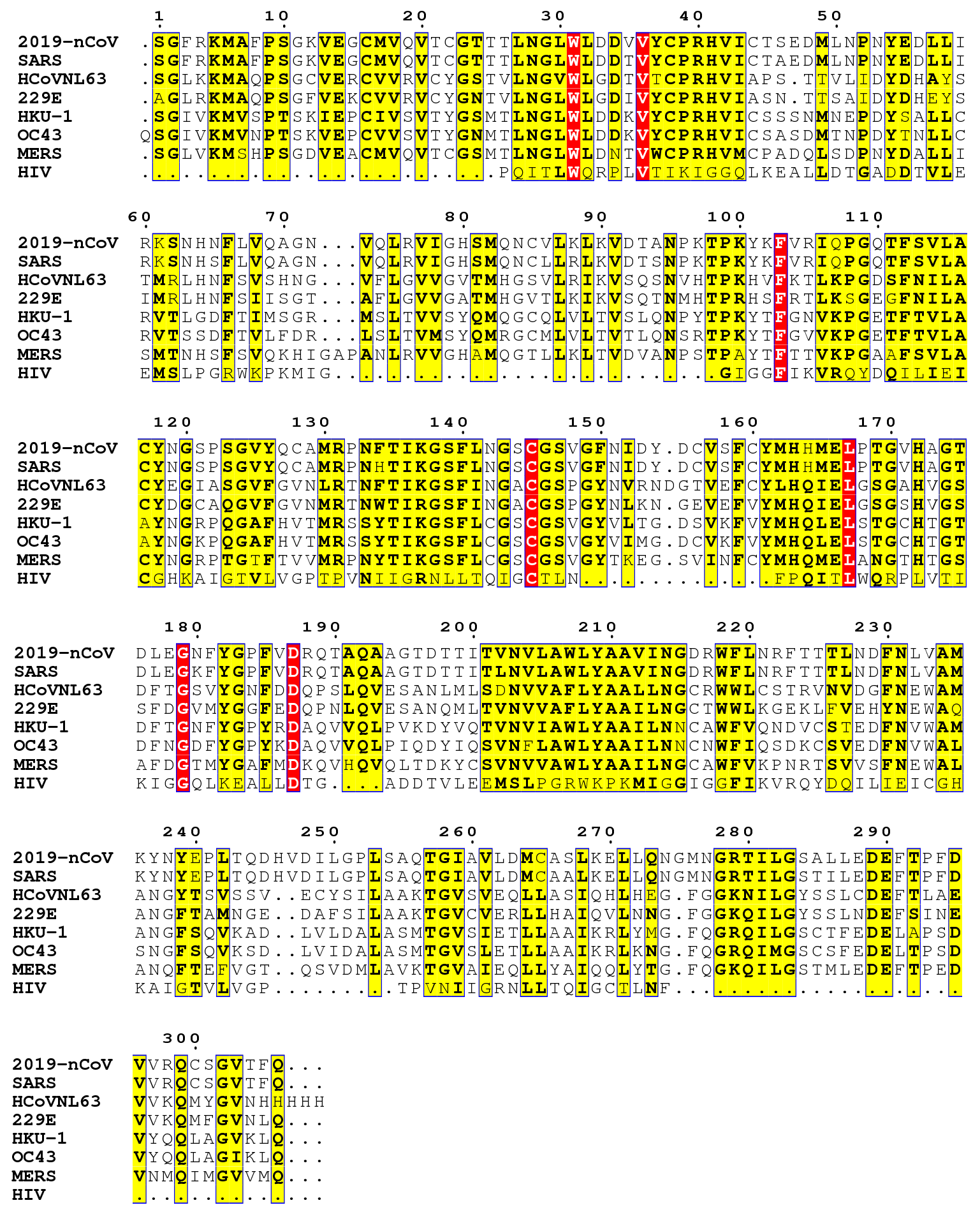}
	\caption{The 3CL protease sequence alignment between SARS-CoV-2, SARS-CoV, MERS, OC43, HCoVNL63, HKU-1, 229E, and HIV.}
	\label{fig:seq-align}
\end{figure}

It is seen that the SARS-CoV-2 3CL protease is very close to the SARS-CoV 3CL protease, but distinguished from other proteases. SARS-CoV-2 has a strong genetic relationship with SARS-CoV, the sequence alignment in Figure \ref{fig:seq-align} further confirms their relationship.

\begin{figure}[ht!]
	\centering
	\includegraphics[scale=0.4]{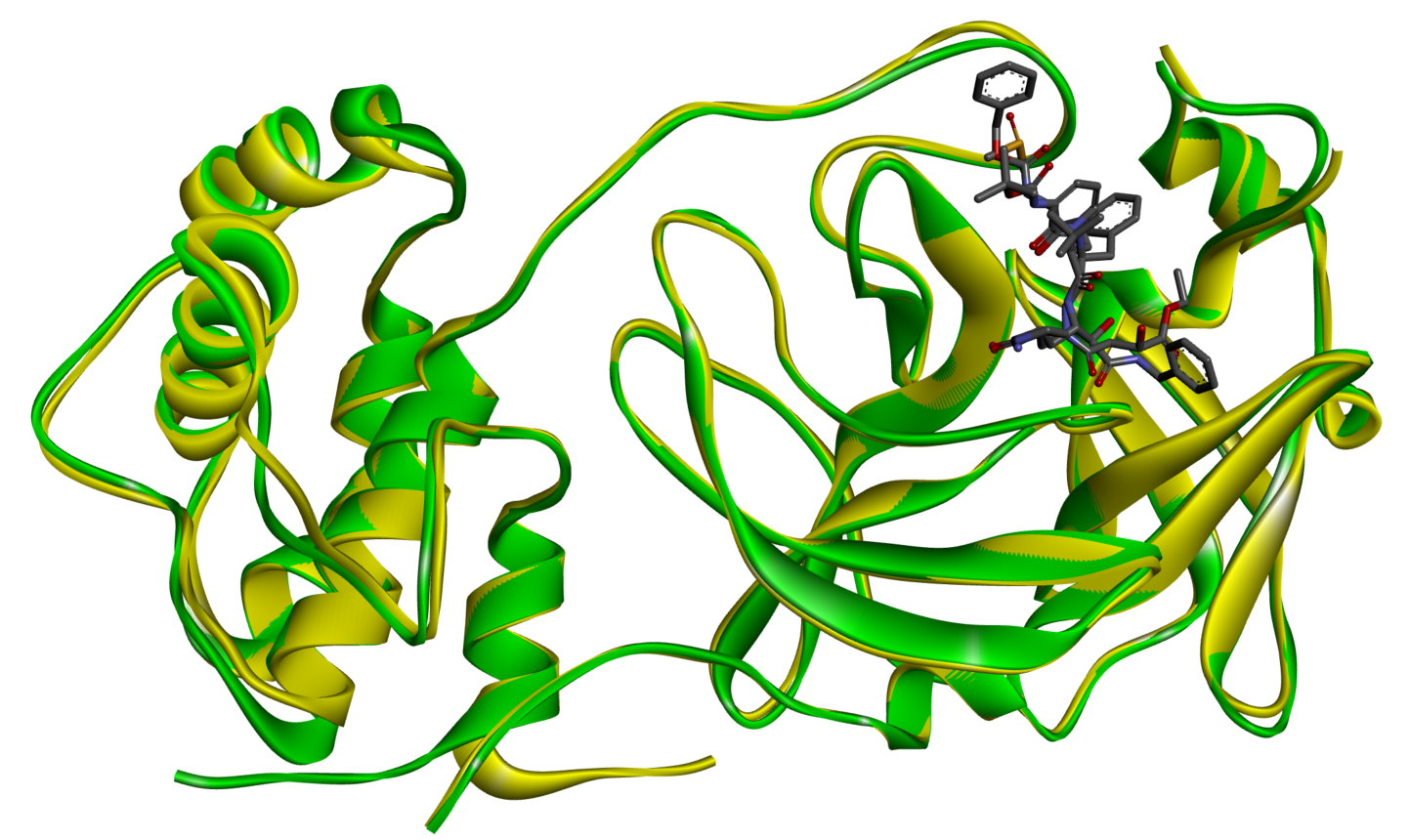}
	\caption{Illustration of the 3D structure similarity between the SARS-CoV-2 3CL protease (PDB ID: 6Y2F, in gold) and SARS-CoV 3CL protease (PDB ID: 2A5I, in green).	The anti-SARS inhibitors in dark color indicate the binding site.
	}
	\label{fig:similarity}
\end{figure}

Not only are the sequences highly identical, but also, as shown in Fig. \ref{fig:similarity} the 3D crystal structures of the SARS-CoV-2 3CL protease is also substantially similar to that of SARS-CoV 3CL protease. Particularly, the RMSD of two structures at the binding site is only $0.42$ \AA.

The high sequence and structure similarity between the two proteases suggests that anti-SARS-CoV chemicals can be equally effective for the treatment of SARS-CoV-2. It means the available experimental data of SARS-CoV protease inhibitors can also be used as the training set to discover new inhibitors of SARS-CoV-2 protease. Our SARS-CoV-2 BA training set contains 314 compounds with their binding affinities to the SARS-CoV or SARS-CoV-2 3CL protease from single-protein experiments available.

 \subsection{The training set of SARS-CoV-2 3CL protease  inhibitors}

We collect the training set from single-protein experimental data of SARS/SARS-CoV-2 3CL protease in public databases or the related literature.

ChEMBL is a manually curated database of bioactive molecules \cite{gaulton2012chembl}. Currently, ChEMBL contains more than 2 million compounds only in the SMILES string format. In ChEMBL, we find 277 SARS-CoV or SARS-CoV-2 3CL protease inhibitors with reported Kd/IC50 from single-protein experiments.

Another database is PDBbind. The PDBbind database includes all the protein-ligand complexes with the crystal structures deposited in the Protein Data Bank (PDB) and their binding affinities in the form of Kd, Ki or IC50 reported in literature \cite{wang2004pdbbind}. The newest PDBbind v2019 consists of 17,679 complexes as well as the binding affinities. We find another 30 inhibitors in the PDBbind v2019.

Additionally, binding affinities for four other SARS-CoV main protease inhibitors and three other SARS-CoV-2 main protease inhibitors are extracted from Ref. \cite{bacha2008development} and Ref. \cite{zhang2020crystal}, respectively. Therefore, we collected 314 SARS-CoV/SARS-CoV-2 3CL protease inhibitors with available experimental binding affinities.

\begin{figure}[!htb]
	\centering
	\includegraphics[width=0.4\textwidth]{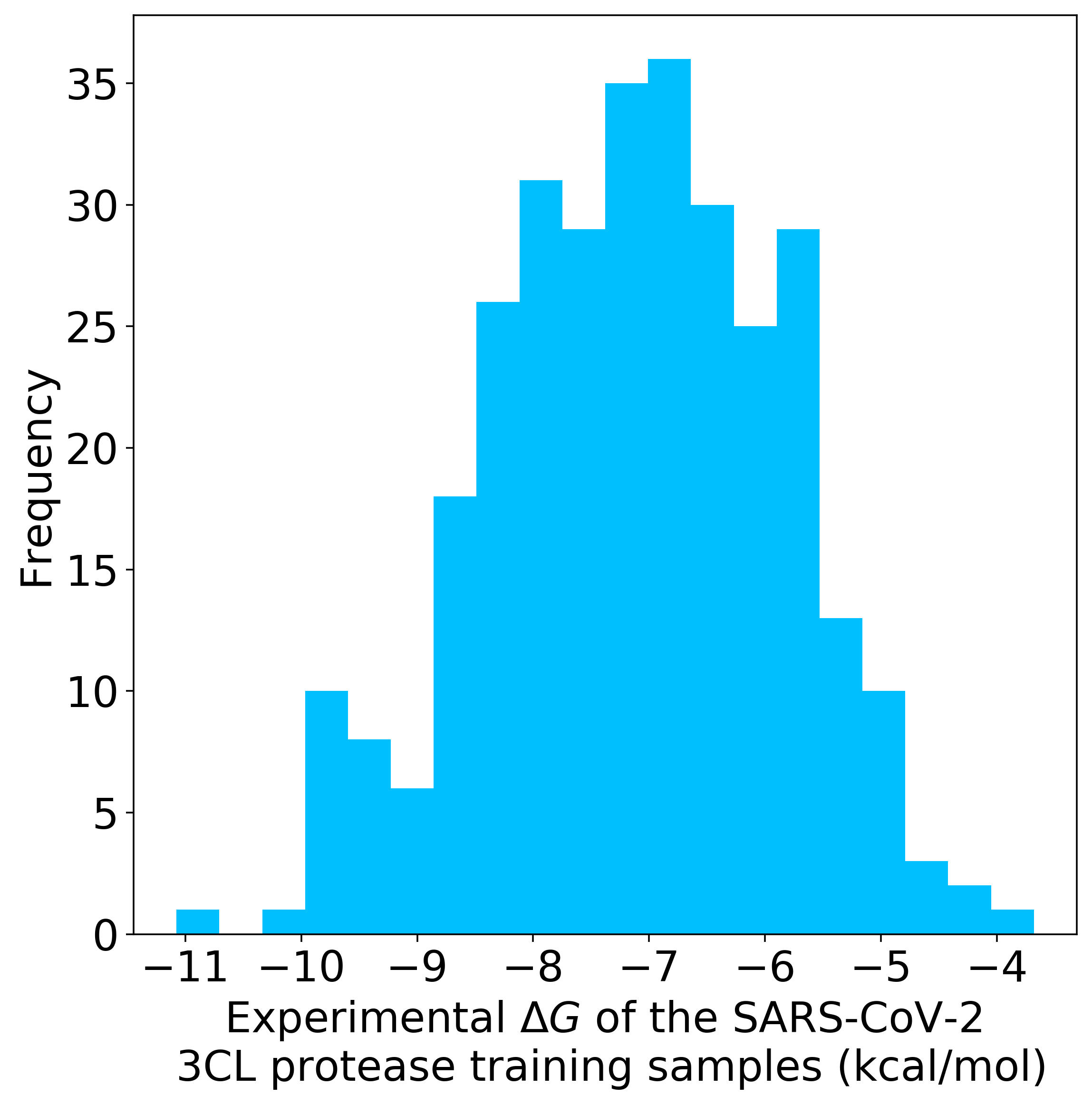}
	\caption{The experimental $\Delta G$ distribution of the training set of SARS-CoV-2 3CL protease  inhibitors.}
	\label{fig:cov2_dist}
\end{figure}
The binding affinity range in this set is from -3.68 kcal/mol to -11.08 kcal/mol. The distribution is depicted in Figure \ref{fig:cov2_dist}. The top 30 inhibitors in the training set are summarized in Table \ref{Summary-trainset}.

\subsection{The FDA-approved drugs and investigational or off-market ones}

DrugBank (www.drugbank.ca) \cite{wishart2018drugbank} is a richly annotated, freely accessible online database that integrates massive drug, drug target, drug action, and drug interaction information about FDA-approved drugs as well as investigational or off-market ones. Due to the high quality and sufficient information contained in, the DrugBank has become one of the most popular reference drug resources used all over the world. In current work, we extract 1553 FDA-approved drugs and 7012 investigational or off-market ones from DrugBank, and evaluate their binding affinities to the SARS-CoV-2 3CL protease.

 \subsection{The log P, log S and synthesizability calculations}

In this work, the log P and synthesizability values are calculated by RDKit \cite{landrum2006rdkit}; the synthesizability in RDkit is in terms of synthetic accessibility score, 1 means the easiest, 10 means the hardest. The log S values are obtained via Alog PS 2.1 \cite{tetko2005virtual}.

\subsection{MathPose}
The 3D binding poses in this work are predicted by the MathPose, a 3D pose predictor which converts SMILES strings into 3D poses with references of target molecules. It was the top performer in D3R Grand Challenge 4 in predicting the poses of 24 beta-secretase 1 (BACE) binders \cite{nguyen2019mathdl}. For one SMILES string, around 1000 3D structures can be generated by a common docking
software tool, i.e., 
GLIDE \cite{friesner2004glide}. Moreover, a selected set of known complexes is re-docked by the three docking software packages mentioned above to generate at 100 decoy complexes per input ligand as a machine learning training set. The machine learning labels will be the calculated root mean squared deviations (RMSDs) between the decoy and native structures for this training set. Furthermore, MathDL models \cite{nguyen2019mathdl} are set up and applied to select the top-ranked pose for the given ligand.

\subsection{The machine learning-based binding affinity predictor}

In the current work, we develop a machine learning model for predicting the binding affinities of SARS-CoV-2 inhibitors. Since the size of the training set in our current case study is only 314, we apply the gradient boosting decision tree (GBDT) \cite{schapire2003boosting} model because of its accuracy for handling small datasets. This GBDT predictor is constructed using the gradient boosting regressor module in scikit-learn (version 0.20.1) \cite{pedregosa2011scikit}.

 The 2D fingerprints of compounds are used as the input features to our GBDT predictor. Our previous study \cite{gao20202d} shows that, the consensus of ECFP4 \cite{rogers2010extended}, Estate1\cite{hall1995electrotopological} and Estate2 \cite{hall1995electrotopological} fingerprints performs the best on binding-affinity prediction tasks. In this work, we  also make use of this consensus. The 2D fingerprints are calculated from SMILES strings using RDKit software (version 2018.09.3) \cite{landrum2006rdkit}. The GBDT parameters in our predictor are, for ECFP4, n\_estimators=10000,max\_depth=7,min\_samples\_split=3,learnin\_rate=0.01,
 subsample=0.3,max\_features='sqrt'; for Estate1 and Estate2, n\_estimators=2000,max\_depth=9,min\_samples\_split=3,
 learnin\_rate=0.01,subsample=0.3,max\_features='sqrt'.

 \subsection{The 10-fold cross-validation of the binding affinity predictor}

 \begin{table}[!htb]
 	\caption{The 10-fold cross-validation test of the machine learning model on the SARS-CoV-2 BA training set.  }
 	\centering

 	\begin{tabular}{lccc|lccc}
 		\toprule
 		& $R_p$ & $\tau$ & RMSE (kcal/mol) & & $R_p$ & $\tau$ & RMSE (kcal/mol)\\\midrule
 		Fold 1 (Train) & 0.998 & 0.976 & 0.089 & Fold 6 (Train) &  0.997 & 0.972 & 0.097 \\
 		Fold 1  (Test) & 0.777 & 0.565 & 0.762 &Fold 6  (Test) &  0.746 & 0.541 & 0.774 \\
 		\midrule
 		Fold 2 (Train) & 0.997 & 0.970 & 0.102 & Fold 7 (Train) & 0.997 & 9.971 & 0.098 \\
 		Fold 2 (Test) & 0.807 & 0.654 & 0.804 & Fold 7  (Test) &  0.812 & 0.636 & 0.822\\
 		\midrule
 		Fold 3 (Train) & 0.997 & 0.971 & 0.096 & Fold 8 (Train) & 0.997 & 0.968 & 0.099 \\
 		Fold 3 (Test)  & 0.822 & 0.632 & 0.743 &Fold 8  (Test) & 0.758 & 0.550 & 0.776 \\
 		\midrule
 		Fold 4 (Train) & 0.997 & 0.969 & 0.101 & Fold 9 (Train) & 0.997 & 0.971 & 0.098 \\
 		Fold 4 (Test) & 0.801 & 0.595 & 0.756 & Fold 9  (Test)  & 0.709 & 0.536 & 0.909\\
 		\midrule
 		Fold 5 (Train) & 0.998 & 0.976 & 0.081 & Fold 10 (Train) & 0.997 & 0.968 & 0.104 \\
 		Fold 5 (Test) & 0.751 & 0.541 & 0.815 & Fold 10 (Test) & 0.792 & 0.620 & 0.799 \\
 		\midrule
 		Average (Train) & 0.997 & 0.971 & 0.097 & & & &  \\
 		Average (Test) &  0.777 & 0.587 & 0.796 & & & & \\
 		\bottomrule
 	\end{tabular}
 	\label{tab:10-fold}
 \end{table}

 In this section, we validate the performance of our machine learning predictor for the 314 inhibitors in the SARS-CoV-2 BA training set with the description provided in Table \ref{tab:10-fold}.

 Table \ref{tab:10-fold} reveals that our machine learning predictor is trained with the averaged train accuracy being the Pearson correlation coefficient ($R_p$) = 0.997, the Kendall's $\tau$ ($\tau$) = 0.971, RMSE = 0.097 kcal/mol. These metrics are based on the averaged values across 10 folds and these results indicate our model is well trained. Their averaged test performances across 10-fold of the whole SARS-CoV-2 BA set are found to be $R_p$ = 0.777, $\tau$ = 0.587, RMSE = 0.796 kcal/mol. These results endorse the reliability of our model in the binding affinity prediction of SARS-CoV-2 inhibitors.

\section{Conclusion}

 The current pneumonia outbreak caused by severe acute respiratory syndrome coronavirus 2 (SARS-CoV-2), has evolved into a global pandemic. Although currently there is no effective antiviral medicine against the SARS-CoV-2, the 3CL proteases of SARS-CoV-2 and SARS-CoV have a sequence identity of 96.1\% and the binding-site RMSD of $0.42$ \AA, which provides a solid basis for us to hypothesize that all potential anti-SARS-CoV chemotherapies are also effective anti-SARS-CoV-2 molecules. In this work, we curate 314 SARS-CoV-2/SARS-CoV 3CL protease inhibitors with available experimental binding data from various sources to form a machine learning training set.
 Using this training set, we develop a gradient boosted decision trees (GBDT) model to predict the binding affinities of potential SARS-CoV-2 3CL protease inhibitors. The 10-fold cross-validation shows our model
 has a Pearson correlation coefficient of 0.78 and a relatively low root mean square error of  0.80 kcal/mol.
 A total of 8565 drugs from DrugBank is evaluated by their predicted binding affinities. We highlight 30 FDA-approved drugs as well as 30 investigational or off-market drugs as potentially potent medications against SARS-CoV-2. We also analyze the druggability of  some potent inhibitors in our training set. This work serves as a foundation for the further experimental development of anti-SARS-CoV-2 drugs.

\section*{Supporting Material}
 The tables of experimental binding affinities for 314 SARS-CoV-2 3CL protease inhibitors, the predicted binding affinities of 1553 FDA-approved drugs and 7012 investigational or off-market drugs are available in Supporting-Material.xlsx. Also, a brief introduction of the MathDL developed by us is in supplementary figures.

\section*{Acknowledgment}
This work was supported in part by NIH grant  GM126189,   NSF Grants DMS-1721024,  DMS-1761320, and IIS1900473,  Michigan Economic Development Corporation,  Bristol-Myers Squibb,  and Pfizer.
The authors thank The IBM TJ Watson Research Center, The COVID-19 High Performance Computing Consortium, and  NVIDIA for computational assistance.
%

\section*{Competing interests}
The author declare no competing interests.


\end{document}